\input harvmac.tex
\input epsf

\ifx\epsfbox\UnDeFiNeD\message{(NO epsf.tex, FIGURES WILL BE IGNORED)}
\def\figin#1{\vskip2in}
\else\message{(FIGURES WILL BE INCLUDED)}\def\figin#1{#1}\fi
\def\tfig#1{{\xdef#1{Fig.\thinspace\the\figno}}
Fig.\thinspace\the\figno \global\advance\figno by1}
%
  
\def\IB{\relax\hbox{$\inbar\kern-.3em{\rm B}$}}
\def\IC{\relax\hbox{$\inbar\kern-.3em{\rm C}$}}
\def\ID{\relax\hbox{$\inbar\kern-.3em{\rm D}$}}
\def\IE{\relax\hbox{$\inbar\kern-.3em{\rm E}$}}
\def\IF{\relax\hbox{$\inbar\kern-.3em{\rm F}$}}
\def\IG{\relax\hbox{$\inbar\kern-.3em{\rm G}$}}
\def\IGa{\relax\hbox{${\rm I}\kern-.18em\Gamma$}}
\def\IH{\relax{\rm I\kern-.18em H}}
\def\IK{\relax{\rm I\kern-.18em K}}
\def\IL{\relax{\rm I\kern-.18em L}}
\def\IP{\relax{\rm I\kern-.18em P}}
\def\IR{\relax{\rm I\kern-.18em R}}
\def\IZ{\relax\ifmmode\mathchoice
{\hbox{\cmss Z\kern-.4em Z}}{\hbox{\cmss Z\kern-.4em Z}}
{\lower.9pt\hbox{\cmsss Z\kern-.4em Z}}
{\lower1.2pt\hbox{\cmsss Z\kern-.4em Z}}\else{\cmss Z\kern-.4emZ}\fi}

\def\CC {{\cal C}}
\def\CD {{\cal D}}

\def\CF {{\cal F}}
\def\CG {{\cal G}}

\def\CN {{\cal N}}
\def\CO {{\cal O}}

\def\CS {{\cal S}}

\def\CV {{\cal V}}

\def\CZ {{\cal Z}}

\def\p{\partial}


\def\zb {\bar{z}}


\def\Tr{{\rm Tr}}

\def\p{\partial}

\def\np{\nabla_{\partial}}

\def\physrev{{\it Phys. Rev. }}

\def\inbar{\,\vrule height1.5ex width.4pt depth0pt}
\font\cmss=cmss10 \font\cmsss=cmss10 at 7pt

\def\a{\alpha}

\def\b{\beta}

\def\e{\epsilon}
\def\m\mu 
\def\n{\nu}
\def\s{\sigma}

\def\p{\partial}
\def\R{\relax{\rm I\kern-.18em R}}
\font\cmss=cmss10 \font\cmsss=cmss10 at 7pt
\def\Z{\relax\ifmmode\mathchoice
{\hbox{\cmss Z\kern-.4em Z}}{\hbox{\cmss Z\kern-.4em Z}}
{\lower.9pt\hbox{\cmsss Z\kern-.4em Z}}
{\lower1.2pt\hbox{\cmsss Z\kern-.4em Z}}\else{\cmss Z\kern-.4em Z}\fi}
\def\pl{{\it  Phys. Lett.}}

\def\np{{\it Nucl. Phys. }}
\def\pr{{\it Phys.Rev. }}

\def\r{{\rm Re}}
\def\i{{\rm Im}}

\def\hf{{1\over 2}}

\def\eps{\epsilon}

\def\la{\lambda}
\def\be{\beta}
\def\ra{\rightarrow}
\def\SYM_4{SYM$_4$}
 %
  %
\def\z{\zeta}
\def\sn{{\rm sn} }

\def\dn{{\rm dn} }


\lref\KKN{V. Kazakov, I. Kostov, N. Nekrasov, ``D-particles, Matrix 
Integrals and KP hierachy'', hep-th/9810035,
\np B (to be published).} 
\lref\HA{M.~Claudson and  M.B.~Halpern, "Supersymmetric Graund State 
Wave Function", 
Nucl. Phys. B 250(1985) 689.}
\lref\Hoppe{J. Goldstone, unpublished; 
J.~Hoppe,  "Quantum theory of a massless
relativistic surface ...", MIT PhD Thesis 1982, and  Elementary 
Particle Research Journal
(Kyoto) 80 (1989).} 
\lref\DeWHoppeN{      
B. de Wit , J. Hoppe, H. Nicolai, "On the Quantum Mechanics of 
Supermembranes", \np B305 (1988)  545.}
  \lref\NEKdiss{N. Nekrassov, {\it PhD Thesis}, Princeton Univ. , 
1997.}
  \lref\NekBau{L. Baulieu, A. Losev and N. Nekrasov, ``Chern-Simons
and Twisted Supersymmetry in Higher Dimensions'', 
 \np B522 (1998)    1998 ; hep-th/9707174.}
  \lref\BFSS{ T. Banks, W. Fischler, S.H. Shenker,
 L. Susskind, ``M Theory As A Matrix Model: A Conjecture'',
\pr D55 (1997)   1997; hep-th/9610043 .} 
  \lref\DK{Michael R. Douglas and Vladimir A. Kazakov, ``Large $N$
Phase Transition in Continuum QCD$_2$'',
\pl B319 (1993) 219 hep-th/9305047. }
  \lref\MNSI{G.~Moore, N.~Nekrasov and S.~Shatashvili, ``Integrating
over Higgs branches'', hep-th/9712241.}
\lref\MNSII{G.~Moore, N.~Nekrasov, S.~Shatashvili,
``D-particle bound states and generalized instantons'', HUTP-98/A008,
ITEP-TH-8/98, hep-th/9803265.}
\lref\GG{M.~Green and  M.~Gutperle, ``$D$-particle bound
states and the $D$-instanton measure'',
hep-th/9711107.}
\lref\WITTEN{E. Witten, ``String theory dynamics
in various dimensions,''
hep-th/9503124, Nucl. Phys. {\bf B} 443 (1995) 85-126.    }
\lref\BUKA{D. Boulatov and V. Kazakov, ``One dimensional string theory
with vortices as an upside down matrix oscillator'', {\it J.Mod.Phys} 
A8 (1993)809; 
D. Gross and I. Klebanov, \np  B359 (1991) 3.}
\lref\KVH{I.K. Kostov and  P. Vanhove, ``Matrix String Partition
Functions'',  \pl B444 (1998)196, 
hep-th/9809130.}

\lref\ikkt{N. Ishibashi, H. Kawai, Y. Kitazawa and A.
Tsuchiya, ``A Large-N Reduced Model as Superstring'', 
Nucl. Phys.  B498 (1997) 467.}
\lref\SAKA{S.
Kakei, ``Toda lattice hierarhy and Zamolodchikov's conjecture",
solv-int/9510006 } 
\lref\rfKS{ W.~Krauth, H. Nicolai and
M.~Staudacher, ``Monte Carlo Approach to M-Theory'', \pl B 431 (1998)
31; W.~Krauth and M.~Staudacher, ``Finite Yang-Mills Integrals'',\pl 
B 435 (1998) 350.}

 
\lref\SEST{S.~Sethi and  M.~Stern, ``$D$-brane bound states
redux,'',   Comm. Math. Phys. 194 (1998) 675,   hep-th/9705046.}
\lref\yi{P.~Yi, ``Witten Index and Threshold Bound States of
D-Branes'', 
  Nucl. Phys. B 505 (1997) 307.}
 
\lref\kato{S. Hirano and  M. Kato, Prog.~Theor.~Phys. 98 (1997) 
1371.}
 \lref\GW{D. Gross and E. Witten, "Possible Third Order Phase 
Transition in the Large $N$  Lattice Gauge Theory", \pr D21 (1980) 
446.  }
\lref\Witdgt{ E.~ Witten,  Commun. Math. Phys.  141  (1991) 153.}
 \lref\Witr{E.~ Witten, ``Introduction to Cohomological Field
Theories", Lectures at Workshop on Topological Methods in Physics,
Trieste, Italy, Jun 11-25, 1990, Int. J. Mod. Phys. {\bf A6} (1991)
2775.}
\lref\Witgrav{E.~ Witten, ``Topological Gravity'',
Phys.Lett.206B:601, 1988.}
\lref\witttft{E.~ Witten, ``Topological Quantum Field Theory",
Commun. Math. Phys. {\bf 117} (1988) 353.}
\lref\Witr{E.~ Witten, ``Introduction to Cohomological Field
Theories", Lectures at Workshop on Topological Methods in Physics,
Trieste, Italy, Jun 11-25, 1990, Int. J. Mod. Phys. {\bf A6} (1991)
2775.}
\lref\mgaudin{M. Gaudin, {  ``Une famille {\`a} un paramettre 
d'ensembles unitaires''}, \np 85 (1966) 545-575 (The text can be found 
in 
the book {\it Travaux de Michel Gaudin}, Les Editions de Physique 
1995)}
\lref\rfBFSS{T.~Banks, W.~Fischler, S.~Shenker, and L.~Susskind, 
\physrev55(1997)5112, hep-th/9610043.}
\lref\FS{{\it B.A.Fuchs,B.V.Shabat,
Functions of a complex variable},
Jawahar Nagar, Delhi; Hindustan Publ.Corp. 1966
}
\lref\BF{Byrd and Friedman,  {\it Handbook of Ellyptic Integrals for 
Engineers and Physicists},  Springer-Verlag, 1954.}
\lref\witloc{E. Witten, "Two Dimensional Gauge Theories Revisited", 
J. Geom. Phys. 9 (1992) 303, hep-th/9204083.}
\lref\sugino{F. Sugino, "Cohomological field theory
approach to matrix strings", hep-th/9904122.}
\lref\KM{ V. Kazakov and A. Migdal, ``Recent Progress in the Theory of
Noncritical Strings", \np B 311 (1988) 171.}



\Title{}
{\vbox{\centerline
 {Dimensionally Reduced SYM$_{4}$ }
\centerline{as   Solvable Matrix Quantum Mechanics}
 \vskip2pt
}}
 \centerline{Jens Hoppe\footnote{$ ^\diamond $}{{\tt 
hoppe@aei-potsdam.mpg.de}}}
\centerline{{\it  Max-Planck-Institut f\"ur Gravitationsphysik,
Albert-Einstein-Institut, D-14476 Golm    }}
\centerline{Vladimir Kazakov \footnote{$ ^\bullet $}{{\tt 
kazakov@physique.ens.fr}}}
\centerline{{\it $^1$  Laboratoire de Physique Th\'eorique de l'Ecole
Normale Sup\'erieure \footnote{$ ^\ast $}{
Unit\'e Mixte du
Centre National de la Recherche Scientifique,
associ\'ee \`a l'Ecole Normale Sup\'erieure et \`a
l'Universit\'e de Paris-Sud.}}}

\centerline{{\ \ \ \it  75231 Paris CEDEX, France}}
\centerline{Ivan K. Kostov \footnote{$ ^\dagger$}{member of 
CNRS}\footnote{$^\circ$}{{\tt kostov@spht.saclay.cea.fr}}}
\centerline{{\it C.E.A. - Saclay, Service de Physique 
Th{\'e}orique }}
 \centerline{{\it 
  F-91191 Gif-Sur-Yvette, France}}

 \vskip 1cm
\baselineskip8pt{
 
\vskip .2in
 
\baselineskip8pt{
We study the quantum mechanical model obtained as a dimensional
reduction of $\CN=1$ super Yang-Mills theory to a periodic light cone
"time".  After mapping  the theory to a cohomological field theory,
the partition function (with periodic boundary conditions) regularized
by a massive term appears to be equal to the partition function of the
twisted matrix oscillator.  We show that this partition function
perturbed by the operator of the holonomy around the time circle is a
tau function of Toda hierarchy. We solve the model in the large $N$
limit and study the universal properties of the solution in the
scaling limit of vanishing perturbation.  We find in this limit a
phase transition of Gross-Witten type.

\bigskip
 \rightline{ SPhT/t99/072}
 \rightline{ LPTENS-99/25}

\Date{July, 1999}  


\newsec{Introduction }


The supersymmetric Yang-Mills (SYM) theories have rich physical
content and their quantitative analysis is in general as difficult as
in the usual, nonsupersymmetric, gauge theories.  However they often
contain, unlike the purely bosonic YM theories, specific sectors,
which can be analysed exactly and where the supersymmetry leads to a
nilpotent (topological) symmetry \WITTEN . The dimensionally reduced
versions of the SYM theory even allow various massive deformations
conserving this symmetry.

In refs. \MNSI\ and \MNSII\ this symmetry (in the zero dimensional
reductions of SYM) was   applied for the calculation of the
(bulk part of the) Witten index for ensembles of $N$ 0-branes in 4, 6
and 10 dimensions,  justifying the conjectures related to the
existence of bound states of zero-branes \GG .  In \KKN\ the method
was applied to study certain correlators of BPS states or, in other
words, of perturbations of the original reduced SYM theories, which
preserve part of the supersymmetry. In the case of the zero
dimensional reduction of $\CN=1$ SYM theory, the large $N$ limit was
studied exactly using the method of \Hoppe\ or by the corresponding
integrability properties allowing to write an explicit (KP)
differential equation for the partition function. One of the
unexpected results was that, in the large $N$ limit, the physical
quantities exhibit an essential singularity at $\lambda =0$, where
$\lambda$ is the coupling of the massive perturbation.  The large $N$
limit of the dimensionally reduced SYM theories is also interesting
because it may reveal part of the structure of the nonreduced
theories, due to the Eguchi-Kawai mechanism.

In this paper we study the one dimensional reduction of the $\CN=1$
$SU(N)$ pure Yang-Mills theory. Unlike the well known and widely used
1d reduction to the usual physical time
 \Hoppe , \HA ,  or
\BFSS, we will retain the "time" along the light cone direction
compactified on a circle of radius $\beta$.  Only this reduction
allows the direct use of Witten's localization principle \WITTEN
\foot{Our argument follows essentially   the construction 
proposed in \refs{\NEKdiss ,\NekBau}, which allows to lift by one the
dimension of the spacetime without loosing the
supersymmetry. Technically speaking, our procedure of dimensional
reduction replaces the Euclidean spacetime by a point and at the same
time introduces the ``time'' dimension $\tau$.  The latter might be
interpreted as a lightlike dimension of the original spacetime, but we
do not know to what extent this interpretation is justified.}.  In
order to get rid of the zero modes of the bosonic fields we will
deform the theory by a massive perturbation corresponding to the
$O(2)$ twisting of the boundary conditions on the time circle with
respect to a subgroup of the euclidean symmetry $O(4)$.  The SYM
theory reduced in this way appears to be identical to the compactified
hermitian matrix oscillator with $SU(N)$-twisted boundary conditions.
The twisting angles are related to a global mode of the (time-like)
gauge field.  This model can be further reduced to that of a unitary
(already time-independent) twist matrix.  We find that the model is
integrable in the sense that its partition function is a tau-function
of Toda hierarchy, i.e. it obeys a chain of nonlinear Toda equations.

The model can be solved exactly and rather explicitly in the large $N$
limit.  The solution of the corresponding saddle point equation and
its physical consequences in the limit of vanishing perturbation
represent the main result of this paper. The solution is parametrized
in terms of elliptic functions.  The analysis of the solution as a
function of the two parameters $\beta\epsilon$ and $\lambda$, where
$\beta$ is the compactification length, $ \epsilon$ is the strength of
the massive perturbation, and $\lambda $ is the twist coupling reveals
the following phenomena:
 
\noindent
1. In the double limit $\e \ra 0$, $\la\ra 0$ the free energy is
universal (under certain deformations) function of the ratio $\e/\la$.

\noindent
2. In the limit of vanishing massive perturbation ($\epsilon \to 0$),
the observables exhibit an essential singularity $\sim\exp \left(-
{{\rm const}\over \epsilon}\right)$.

\noindent
3. The analytical continuation of the model at the point
$\e\beta=i\pi$ (inverted oscillator) shows the scaling of the $c=1$
compactified noncritical string theory. This value of the
compactification length $\e\beta$ corresponds to the
Kosterlitz-Thouless critical point (see appendix C).

The paper is organized as follows.  In section 2 we describe the
reduction of the partition function of the one-dimensionally reduced
$\CN=1$ SYM$_4$ theory to that of the reduced twisted matrix
oscillator, by the use of the supersymmetry and the localization
theorem.  We then reduce the configuration space of the model to the
set of the eigenvalues of the unitary twist matrix.  In section 3 we
find that the partition function of our model is a $\tau$ function of
the Toda integrable hierarchy and write the differential equations
satisfied by the partition grand canonical function.  In section 4 we
give an exact solution of the saddle point equation for the large $N$
limit of the model in terms of elliptic parametrization; the
calculations are presented in Appendix A.  In section 5 we study the
limit of small massive perturbation.  We find an universal expression
for the free energy in presence of a source for the Wilson loops, in
the scaling limit $\epsilon\to 0$ and $\la\to 0$.  We analyse the
properties of the solution, especially in the small compactification
radius limit and near the curve of the Gross-Witten type transitions.
Section 6 is devoted to conclusions.  In Appendix C we give the
solution of the analytic continuation of our model to imaginary time
$\e\beta = i\pi$, which is the Kosterlitz-Thouless point for the
corresponding $c=1$ noncritical string.

\newsec{  Definition of the model and its reduction to 
one-dimensional matrix quantum mechanics}

In this section we will show that the dimensionally reduced $\CN=1$
super Yang-Mills theory with gauge group $SU(N)$ can be mapped to
one-dimensional matrix quantum mechanics. The dimensional reduction
consists in replacing the $4$-dimensional (Euclidean) space-time by a
single lightlike "time".
 
Let us first give the generalization of the argument of \KKN\ to the
case of one dimension.  We start with the the $\CN=1$ $SU(N)$ SYM$_4$
containing $4$ bosonic matrix fields $A_\mu \ \ (\mu = 0,1,2,3)$, and
4 real fermionic fields $\Psi_{\a}\ \ (\a = 1, \ldots, 4)$.  After
performing a Wick rotation $x_{0} = -i x_{4}$, the action of the
Euclidean theory can be written as %
\eqn\action{ \CS=   \int d^4x  {\Tr}\left( -{1\over 
4}F_{\mu\nu}  ^{2}  +{1\over 2}\ \Psi^{T}(\nabla_{4}+ \vec \gamma 
\cdot \vec \nabla)\Psi\right), }
where $\nabla_\mu =i\p_\mu +A_\mu $ is the covariant derivative and
the gamma-matrices are represented as direct products of Pauli
matrices: $\gamma_{i} = \s_{i}\times \s_{i} \ \ (i=1,2,3)$.  The gauge
group $SU(N)$ acts to all fields in the adjoint representation.  
let us assume that all fields depend only on the time-like coordinate
\eqn\taudef{\tau=x_3-ix_4, } which parametrizes a circle with radius
$\b$.  The resulting model is a matrix quantum mechanics containing
four bosonic and four fermionic matrix variables.

We will evaluate the functional integral for this one-dimensional
matrix model by mapping it to a cohomological field theory, which will
allow to apply Witten's localization argument \witloc.  Let us
redefine the fields as
\eqn\redeff{\left(
\matrix{A_{1}\cr
A_{2}\cr A_{3}+i A_{4}\cr A_{3}-i A_{4}\cr}\right)
= \left(
\matrix{X_{1}\cr
X_{2}\cr \phi\cr\bar\phi\cr}\right), \qquad
\Psi = \left(\matrix{\psi_{1}\cr
\psi_{2}\cr
 \eta\cr \chi\cr}\right) } Then the action
\action \ can be written as a BRST  exact form.
The BRST transformation $Q$ acts on  the complex 
of fields $\Phi=\{  X_a, \phi, \bar\phi, H; \psi_a, \eta, \chi
\}$ (where   $H=i[X_1, X_2]$ is considered as an auxilliary field )
as
\eqn\nilp{\eqalign{Q X_{a} = \psi_{a},
\quad & \quad Q \psi_{a} = [i\p_\tau+ \phi, X_{\alpha}]\quad\quad
(a=1,2)
 \cr
Q \bar\phi = \eta, \quad & \quad Q \eta = 
[i\p_\tau +\phi, \bar\phi]
\cr
Q\chi = H, \quad & \quad Q H = [i\p_\tau + \phi, \chi] \cr
\qquad \qquad Q \phi & = 0. \qquad \qquad \cr}}
Namely, the action
$$\eqalign{\CS&= 
 \int_0^\beta d\tau \Tr \Big( i H [ X_{1}, X_{2}]
 + {\half} H^2 +
[X_{a}, i\p_\tau+\phi] [X_{a} , \bar\phi] + {\half}
[i\p_\tau +\phi, \bar\phi]^{2} \cr &+  
\hf \chi \e^{ab} [X_a \psi_b] 
+ [X_a,\eta] \psi_a - 
[\bar \phi, X_a] [i\p_\tau+\phi, X_a] 
+\hf \chi[i\p_\tau+  \phi,\chi] + [\psi_a, \bar \phi] \psi_a\Big) 
\cr}$$
can be written as
\eqn\actred{\eqalign{  
S\ \  &  =Q \int_{0}^{\b} d\tau  \Tr \CV(\Phi)\cr
 \CV(\Phi)&= 
    {\half}
\eta [i\p_\tau +\phi, \bar\phi] + \chi 
(H-i[X_1,X_2])   + \sum_{a=1}^{2} \psi_{a} [X_{a}, \bar\phi]   .\cr}}
The square $Q^2$ of this transformation represents the "time" 
gauge
transformation generated by $\phi$, $\nabla_\tau =
[i\p_\tau+\phi, \ \ ]$.
Hence $ Q$ is nilpotent
on the gauge-invariant quantities. The ghost number of
the fields is $-2$ for $\bar\phi$, $-1$ for $\eta$ and $\chi$, 0 for 
$X_a $ and $H$, $+1$ for $\psi_a$, and $+2$ for $\phi$.

The functional integral with respect to the BRST complex 
of fields $\Phi $ 
\eqn\prtFc{ \CZ_N(\b, g) = \int {\CD \Phi \over {\rm Vol(\CG) }}  
e^{-{1\over g}
\CS[\Phi]}}
(where the integration measure is 
normalized by the 
volume of the gauge group $\CG$)
can be therefore evaluated using the Witten's
localization argument \witloc .
Namely, the integral is saturated by the
BRST critical points  $Q\Phi =0$.
More strictly, we have to integrate 
 over a continuous critical manifold, because of the zero modes of 
$\phi, \chi, \bar\phi$.
The zero modes are elliminated by adding, 
 following \witloc , a $Q$-exact term to the action
\actred\ by changing the  action 
to  $\CS + \delta \CS$, with 
 \eqn\potV{  \delta \CS =  t Q  \int_0^\beta d\tau \Tr \chi \bar\phi.}
  We can also   discard from the very beginning, 
the term   ${\half}
\eta [i\p_\tau +\phi, \bar\phi] $.  
As before, $H$ can be integrated out by setting   
$H= t\bar\phi +[X_1, X_2]$ in \actred. The advantage of introducing 
 the perturbation \potV\ is that for $t\ne 0$, the fields
$\bar\phi, \chi$, and $\eta$ can be integrated out. 
However, the perturbed integral does not coinside
in general with the original one because of 
the new fixed points "flowing in from the infinity "
when one perturbes to $t\ne 0$ \witloc . The correct statement is 
that there exists
a class of  BRST-invariant operators whose 
(nonnormalized) expectation values coincide in the
original and the perturbed theory.
The first such operator
 is
\eqn\pertu{  \omega   = \int_0^\b d\tau 
\Tr (- [i\p_\tau +\phi, X_1] X_2  +\psi_1\psi_2)
=  Q \int_0^\b d\tau \e^{ab}  \Tr  (\psi_a X_b ).
 }
The second is  any $SU(N)$-invariant function 
$f(\Omega) $ of 
the holonomy factor around the circle
\eqn\Witlp{ \Omega = 
   \hat T \exp \left(i \int_0^\b d\tau \phi(\tau)\right).}
 The functional integral for the expectation value 
\eqn\expV{ \langle e^\omega f(\Omega) \rangle
=  \int {\CD \Phi \over {\rm Vol(\CG) }}  e^{-{1\over g}
(\CS +\delta\CS) +\omega }  f(\Omega)}
does not depend on the coupling $t$. Indeed, taking
 the derivative in $t$ and integrating by parts, we find zero, since 
the integrand
vanishes at infinity due to the factor $e^\omega$.
Therefore we can take the limit $t\to\infty$, after which the 
integral \expV\  
 gets localized near
the zeros of $H, \bar\phi, \chi, \eta$.
(In particular, the partition function does not depend on the gauge
coupling $g$.)
Now we can calculate the partition function \prtFc\ 
understood as the average of the 
identity operator.  Since $\omega$ has ghost number $+2$, 
due to the ghost number conservation
\eqn\prtO{\CZ_N(\b , g) \equiv \langle   1 \rangle=  \langle  
e^\omega   \rangle . }
The above  argument has been applied recently by
F. Sugino \sugino\ 
in order to calculate the
partition  function of the  four-dimensional
$\CN=1$  SYM  reduced to a two-dimensional torus,
with periodic boundary  conditions for all fields.
 Our case  is slightly more subtle, because of 
 the zero modes of the fields $X_a$.
These zero modes will be elliminated,
     as in \MNSII\ and later in \KKN , namely by deforming the BRST 
operator in the definition of the 
action    \actred.  
Let us first notice that after the redefinition of the fields the 
theory  is still invariant under
the   $O(2)$ rotations
in the directions orthogonal
to the light cone: 
$$ X_{1} + i X_{2} \ra e^{i\e} \left( X_{1} +i
X_{2} \right), \ \ \psi_{1} + i \psi_{2} \ra e^{i\e} \left( \psi_{1}
+i \psi_{2} \right).$$ 
This allows to construct another BRST operator, which 
squares to a linear combination of a gauge 
transformation and an $O(2)$ rotation.
The twisted   BRST charge $Q_\e$ acts as
\eqn\nilpe{\eqalign{Q_\e X_{\alpha} = \psi_{\alpha},
\quad & \quad
Q_\e \psi_{\alpha} = [i\p_\tau+ \phi, X_{\alpha}]+ 
i \e \varepsilon^{\alpha\beta} X_{\beta},  \cr
Q_\e \bar\phi = \eta, \quad & \quad Q_\e \eta = 
[i\p_\tau +\phi, \bar\phi]
\cr
Q_\e \chi = H, \quad & \quad Q_\e H = [i\p_\tau + \phi, 
\chi] \cr
\qquad \qquad Q_\e \phi & = 0. \qquad \qquad \cr}}
The modification of the supercharge 
is equivalent to changing the action 
\actred\ and the operator  \pertu\ as
 \eqn\pert{\eqalign{
 \CS& \to\CS+ 
  2i \epsilon \int d\tau \Tr(\bar\phi [X_1,X_2]) \cr
  \omega & \to \omega - {\e\over 2} \int_0^\b d\tau 
\Tr (X_1^2+X_2^2).\cr}}   
In the limit $t\to\infty  $  the integral \mit\ gets localized near
the zeros of $H, \bar\phi, \chi, \eta$,  leaving 
the place
to the  action
\eqn\newac{   S=\int d\tau   {\Tr} \left( 
-{i}[i\p_\tau + \phi,X_{1}] X_{2} - {\half} \epsilon ( X^{2}_{1} +
X_{2}^{2}) + \psi_{1} \psi_{2} \right),} 
  and the $\psi$'s can then be integrated out. 
Finally, the integration
over $X_2$ gives the partition function of the matrix oscillator (with
the coordinate $X_1\equiv X$)  in presence of  the one-dimensional 
gauge
 field  $\phi(\tau)$
\eqn\twosc{   \CZ_{N}(\b , g,  \e) =\int {{\CD \phi(\tau)
 } \CD X(\tau) \over{\rm Vol}  \CG  }\exp \left(- \hf \Tr 
\int_0^\b
d\tau \left({1\over \e}[ i\p_\tau + \phi,X]^2+\e X^2  
\right)\right),   }
 with  periodic boundary conditions
$  X(\b)=X(0), \ \ \ \phi(\b)=\phi(0)  . $
  It is clear that the integral depends on $\b$ and $\e$ 
only through the product $\e\b$. We will  absorb 
$\e$ in $\b$,
\eqn\epsbta{ \e \b \rightarrow   \b,}
 remembering that the perturbation is 
lifted in the limit $\b\to 0$.

The functional integral over the field $\phi$ 
can be written, after fixing a gauge $\p_\tau \phi =0$,
as an integral   over the unitary matrix\foot{It is
assumed that the integration contour for the
eigenvalues of $\phi$ is chosen along the real axis.
In this case $\bar\phi$
should be taken anti-hermitian,
see the the discussion in \witloc.} representing
holonomy  factor  defined by \Witlp ,
namely $\Omega= e^{i\b\phi}$
normalized by the volume of $U(N)$.  
The holonomy factor enters the functional integral   
  over $X$ as the    twisted  boundary condition:
\eqn\tbc{ X(\b)=\Omega^+X(0)\Omega. } 
 The integral over $X$ can be performed exactly, and the integral 
over the
unitary matrix $\Omega$ reduces to an integral over its eigenvalues
$e^{i\theta_1}, ..., e^{i\theta_N} $ (which are defined up to a
permutation, hence a combinatorial factor $1/N!$).  The partition
function is therefore given by the $N$-fold integral\foot{ This
happens to be exactly the partition function of the one-dimensional
gas studied by Michel Gaudin in 1966 \mgaudin ; it was extensively
used in \BUKA\ to study the compactified  1+1 dimensional string
theory via   matrix quantum mechanics; in relation to the
actual SYM theory this formula was communicated to us by N. Nekrasov.}
\eqn\twpa{ \CZ_N(\beta )= {1\over N!} 
\oint \prod_{k=1}^N {d\theta_k \over 2\pi} \  
{  \prod_{i\ne j}\sin[\hf ( \theta_i - \theta_j)]
\over \prod_{i,j}\sin [\hf ( \theta_i - \theta_j +i  \beta  )]}}
where $\theta_{i }= \b\phi_{i}$.

 The expectation value we are calculating is
a deformation of the Witten index %
\eqn\pf{  \CZ_N (\b) = \Tr    
  (-)^{F} e^{-\b  H } e^{i\b\e J}  ,}
  where the $(-)^{F}$-factor is included in order to impose the 
periodic 
boundary conditions on the fermionic fields,
 and  the trace is  twisted  by an $O(2)$ rotation $e^{i\b\e J}  $
in  the $(12)$ plane.
 The twisting of the  BRST charge  $Q\to Q_\e$ does not 
change locally the functional integral, but it does
change the boundary conditions for the fields.

Now we are at the most subtle point of the reduction procedure, which
deserves to be discussed in more detail. Considering the fields
$\phi$ and $\bar\phi$ as two independent fields imply that the
integration over them is understood as contour integration.  The
twisting separates the poles and zeroes of the integrand and allows to
evaluate the integral by the residue theorem, in complete similarity
with the calculation of \MNSII\ for the zero dimensional model.  Since
the integrand does not depend on the variable
$$\bar \theta ={ \theta_1+...+\theta_N\over N},$$
the contour integral with respect to this variable would give zero. 
In fact, the integration with respect to this variable should be 
excluded
because this is one of the normalizable zero modes of the 
original fields and the measure $\CD\Phi$ should contain
a product of delta functions of the bosonic and fermionic zero modes.
In particular, the normalized zero mode of $\phi$ is
$$\phi^{(0)}= {1\over \sqrt{ N\beta} } \int_0^\b d\tau \Tr \phi(\tau)
= \sqrt{N\b}\  \bar\theta.$$
For a more detailed discussion see \KVH .        Therefore the 
measure in \twpa\ contains a delta function
$ \delta(\phi^{(0)})\sim  \delta(\theta_1+...+\theta_N)$,
which suppresses the contour integration with respect to $\bar\theta$.
The integral over $\theta$'s is normalized by the volume of the 
residual 
global gauge group.  The introduction of the delta function 
should respect this normalization. Thus we have to insert
$$2\pi\delta (\bar\theta) = 2\pi N \delta(\theta_1+...+\theta_N).$$
Now we can integrate, after representing the integrand as a
determinant using the Cauchy identity, by using the residues theorem.
The integral is equal to the sum of the identical contributions of the
$(N-1)!$ cyclic permutations in the expansion of the determinant 
 \eqn\ptFK{\eqalign{
\CZ_N(  \beta  )&=    (-)^{N-1} {(N-1)! \over N!}
\oint {d\theta_1\over 2\pi}... {d\theta_N\over 2\pi}  2\pi N
\delta(\theta_{1}+ \ldots \theta_{N} ) 
\prod_{k=1}^N {1    \over   \sin [\hf ( \theta_i - \theta_ 
{i-1}+i   \beta)]}\ \cr 
& =  \int {{d\theta_{1} \over 2\pi}\ {d\theta_{N} \over 2\pi}\ 
\delta(  N {\theta_{1}+\theta_{N}\over 2})
  \over \sinh\hf [\theta_{1 }-\theta_{N}- i(N-1)\beta ]  \   
\sinh\hf [\theta_{1 }-\theta_{N} +i \beta]}
  \cr
&={1\over 2N  \sinh N{  \b\over 2} }.\cr}}
In the limit $\b\to\infty$  our  partition  vanishes, which is not 
unexpected,
 since the Witten index of the $D=4$ theory is zero. 
   In the limit  $\b\to 0$ we recover the result for the completely 
reduced theory
$  \sim 1/N^2$,
in agreement with \MNSII. 
A more careful analysis   allows to reproduce also the 
numerical coefficient, in accordance with the conjecture
made in  \rfKS .   
In the limit $\b\to 0$,  the $\Omega$-integral  
is saturated by the integration in the vicinity of the
$N$ central elements of $SU(N)$ which are parametrized by 
the element of the $su(N)$ Lie algebra \SEST.   
  After performing carefully the limit, one finds (see e.g.  \KVH)
$$\CZ_N(\b)\to { ( g/\b)^{{1\over 2}(N^2-1)}
\over \b\ \CF_N} \  \CZ_N^{(0)}\left({g\over\b}\right) $$
where $ \CZ_N^{(0)}( g )$ is the partition function of the 
completely reduced theory, and one reproduces the result
of \MNSII :
\eqn\zNORM{ \CZ_N^{(0)}\left({g}\right)= \CF_N \ g^{-{1\over 2}(N^2-1)} \ {1\over
N^2}.}
 (The numerical factor $\CF_{N}$ depends on the way the
integration measure is normalized. In the normalization used in
\MNSII\ this factor is equal to one, but this is not the most natural
choice from the point of view of applications to the D-brane physics.)
In the particular case of the $SU(2)$ theory the results was
obtaind by the direct calculation of the integrals \yi .
 
Let us  note that our
partition function only formally coincides with that of the 
twisted  matrix oscillator and, at least for finite $N$, 
there is an ambiguity related to the prescription for
the contour integration.
  Witten's localization procedure leads to an integral over the 
{\it Lie algebra}  and   logically 
  the integration 
with respect to $\theta$'s  should be taken  along the whole real 
axis. 
With this definition, only 
$(N-1)!$ terms in the expansion of the determinant will
contribute to it. On the other hand, had we integrate in  interval 
 $[0, 2\pi]$, this would correspond to contour integration with 
respect
 to the eigenvalues of the {\it Lie group}  element
 $\{ t_{k}=e^{i\theta_{k}} \}_{k=1}^{N}$, where the contours circle 
the 
 origin.  In this case we   would get contributions from all 
 $N!$ terms in the expansion of the Cauchy determinant.
The result  would be given, instead of  \ptFK , by (see, for example, 
ref. \BUKA)
\eqn\twOs{
\tilde \CZ_N (\b) = 
{ e^{-N^2\b/2}\over (1-e^{-\b})(1-e^{-2\b})...
 (1-e^{-N\b})}.}
 Unlike \ptFK, 
the $\beta \ra 0$ limit of this formula does
not match  the result of \MNSII .

Which of the two formulas \ptFK\ or \twOs\ is correct? Clearly the
difference between them is due to a different treatment of the
boundary conditions for the field $\phi$ in the formula \twosc .
result
 %
considered
then at
hermitean 
oscillators). 
A happy resolution of this paradox would be that from the point of
view of the application of Witten's localisation principle both
formulas seem to be possible but the result depends on the boundary
conditions and the contours of integration for the field $\phi(t)$ in 
the
original action \actred . However we feel that the question is rather
subtle and more study is needed to clarify it. For example we
cannot be sure that the supersymmetry of the original model isn't
violated in one of two cases. On the other hand, the local BRST 
symmetry
used for the calculations is certainly intact.

Now let us consider a slightly more ambitious problem, namely to 
calculate
the generating functional of a set of BRST invariant  operators
made out of the  gauge field $\phi$.  As mentioned before, such 
operators can be 
constructed as traces of the holonomy  $\Omega$
in different representations, or, equivalently, 
as polynomials of the  moments   $\Tr(\Omega)^k$.
   We will add to the action the simplest possible source term
\eqn\poto{\lambda \Tr(\Omega^+ + \Omega).}  
Repeating the   arguments, which led to 
\twpa , we find for the generating functional the following
integral representation
 \eqn\twpa{ \CZ_N(\beta,\lambda)= {1\over N!} 
\oint \prod_{k=1}^N {d\theta_k \over 2\pi} \   e^{N\lambda
\cos \theta_k}\ \prod_{i\ne j} {\sin [\hf ( \theta_i - \theta_j)]
\over \sin [\hf ( \theta_i - \theta_j +i  \beta )]}}
where $\theta_{i }= \b\phi_{i}$.
If one follows the recipe of \MNSII ,
the integration should be considered as 
a contour integration along the real axis, where the $N$
integration variables are  subjected to the constraint 
$\theta_1+...+\theta_N=0$. Then the  result should 
be analytic as a function of $\beta$, which can therefore
 be given  complex values.
It is plausible that in the large $N$ limit, which we are interested in,
 if the perturbation is sufficiently strong,
the choice of the contours should be  not important. The equivalence
 between  the reduces SYM theory and the twisted matrix oscillator 
should takes place only in this limit.

It would be very interesting to understand
 what is the meaning, in terms of the original   supersymmetric theory \action, of the deformation 
that leads to the partition function \poto. 

The reduction from 4 to 1 dimension of the original theory turns  three of the 
components of the gauge field 
into Higgs fields ($X_1,X_2$ and $\bar \phi$).
This makes the direct calculation of the 
partition function (which  is related to the bulk part of the Witten index) more delicate,
 because of the  absence of mass gap. 
By introducing the deformations \potV ,
\pertu\ and \pert\
we add an additional  Higgs potential, thus 
 breaking part of the supersymmetry.  
  The effect of the source term, which we added to obtain 
the partition function \poto, 
  depends substantially on the way we have perturbed the theory. Indeed, it has positive ghost charge, and its effect would be zero, if
the perturbation \pert\ 
of ghost charge $-2$  were not there to compensate it.  This is also true in the completely 
reduced theory, discussed in ref. \KKN .



\newsec{The partition function as a tau-function of the Toda 
hierarchy}
\def\rh{e ^{\b /2} }
\def\hr{e^{-\b/2}}

Here we will show that our partition function 
with a source term $\lambda \Tr(\Omega^+ + \Omega)$  is a 
tau-function of  discrete Toda 
chain.
Let us rewrite  the partition function
of the model eq.\twpa\  in the following form:
\eqn\Npart{ \CZ_N(\b ,t) = {1\over N!}\oint\prod_{j=1}^N {d z_j\over
2\pi }e^{U(z_j)}{\Delta^2(z)\over \prod_{k, m}
(\rh z_m-\hr z_k)}  }
where   $z_k=e^{i\theta_k}$, $U(z)=\sum_{n\ne
0}t_n z^n$, and $\Delta(z)$ is the Van-der-Monde determinant of 
$z$'s. 
In our case $t_1=t_{-1}=N\lambda $ and 
 $t_{n}=0$  for $n \ne \pm 1$, but
most of the following conclusions are true for a general $U(z)$.

Let us now introduce the grand canonical partition function with the
"charge" $l$:
\eqn\Mpart{ \tilde \tau_l[t,\mu ]=\sum_{N=1}^\infty\ e^{\mu N}
 e^{-l N\b}\CZ_N(\b,t)  }
Due to the Cauchy identity the last equation can be rewritten in terms
of a functional Fredholm determinant:
\eqn\Zdet{\eqalign{
\tilde\tau _l[t,\mu ]& =\sum_{N=1}^\infty e^{\mu N}\ 
{{e^{-l N\b}\over N!}} \oint\prod_{j=1}^N 
{d z_j\over 2\pi }e^{U(z_j)} \ \det_{k,m} \ {1\over 
\rh z_m-\hr z_k} \cr
&=
{\rm Det} (1+e^{\mu-\b l}  \hat K) , \cr}}
where the operator $\hat K$ is defined as 
$$(\hat  Kf)(z)=\oint {dz\over 2\pi}\ {e ^{\hf[U(z)+U(z')]}\over 
\rh z-\hr z'}\ f(z'). $$
It is convenient to modify slightly the definition of the 
tau-function:
\eqn\modt{  \tau_l[T,\mu ]=\tilde\tau_l[t,\mu 
]\exp(-\sum_{n>0}nt_nt_{-n}) ,
}
where  we introduced new couplings $T_n$ by:
\eqn\Tpot{  U(z)=\sum_{n\ne 0} 
z^n T_n(e^{-n\b /2}-e^{ n\b /2}) } 
so that the old couplings are expressed through the new ones as:
\eqn\ONc{  t_n=T_n (e^{-n\b /2}-e^{n\b /2}) . }
We also note that 
\eqn\Tred{  \tau_l[T,\mu ]= \tau_0[T,\mu -\beta l]\equiv  \tau[T,\mu 
-\beta l] .}
Using, for example, the general construction of the paper \SAKA \ for 
our particular tau-function we conclude that it
is a particular case of the tau-function of  Toda hierarchy. It
satisfies the Toda chain equations. Namely let us introduce a new 
function 
\eqn\ePhi{  e ^{\Phi_l}={ \tau_{l-1} \over  \tau_l }=
{ \tau[T,\mu -\beta (l-1)] \over  \tau[T,\mu -\beta l] }  }
and the notations
$$ix_\pm = \sqrt{2}T_{\pm 1}=\pm (\rh-\hr)^{-1} t_{\pm 1}.$$
The first equation of the Toda hierarchy can be written as
\eqn\Toda{ {\p\over\p x_+}{\p\over\p x_-} \Phi_l+
\hf\left( e ^{\Phi_l-\Phi_{l+1}}- e ^{\Phi_{l-1}-\Phi_l}\right)=0     
}

Due to the symmetry $\theta \ra -\theta$ of the measure our
tau-function depends only on the variable $x=\sqrt{x_+x_-}$. The
corresponding reduced equation is:
\eqn\TodaR{ \Phi_l''+{1\over x}\Phi_l'+
\hf\left( e ^{\Phi_l-\Phi_{l+1}}- e ^{\Phi_{l-1}-\Phi_l}\right)=0 
,    
}
where the derivatives are taken with respect to $x$.

For the function $\psi_l=\Phi_l-\Phi_{l+1}=\log{ \tau_{l-1}\tau_{l+1}
\over  \tau_l^2}$ 
the Toda equation reads:
\eqn\TodaP{  \psi_l''+{1\over x}\psi_l'+
\hf\left( 2e ^{\psi_l}- e ^{\psi_{l-1}}- 
e ^{\psi_{l+1}}\right)=0     }

The tau-function, as well as $\Phi_l(0)$ and $\psi_l(0)$,   can
be  determined for  $x=0$ using the methods of
\BUKA\  (for $x=0$ the tau-function  is the grand canonical
partition function of the matrix oscillator in the singlet
representation of the $U(N)$ group, which is the same as the
partition function of $N$ fermionic oscillators) and it can serve as a
boundary condition for the Toda chain equation. For example, one 
finds from \twOs
\eqn\Boun{ e^{\psi_l(0)}={1+e^{-\b(\mu -l-3/2)} \over 1+ e^{-\b(\mu 
-l-1/2)}} . }
Let us also note that $\psi_{l}(x)$ is analytic in $x^{2}$ at the 
origin, which gives the second initial condition 
$\p_{x}\psi_{l}|_{x=0}=0$.

Using these equations and the boundary conditions we expand the
partition function \twpa\  in powers of $\lambda^2$. In the first 
order:

\eqn\frtp{ {1\over 2 N^2}{\p^2 \over \p \la^2} \log \CZ = 
{1-e^{-N\be}\over
1-e^{-\be}}}

This is the simplest correlation function 
$\langle \tr \Omega^+\tr\Omega \rangle$ of the holonomy wilson loop
in our original model.

The large $N$ limit of the initial partition function \twpa\ or 
\Npart\
can be studied in terms of a special scaling limit of these Toda
equations (since $\mu \sim N$ in the  legendre transform
from canonical to microcanonical partition function), similar to the
KP-hierarchy approach of a simpler zero-dimensional model of paper 
\KKN. 
We leave this study to a future publication.

\newsec{Saddle point  equations in the large $N$ limit}

In this section we will investigate the large $N$ limit, which is the 
most interesting from the point of view of
applications.
Since the potential $\lambda\cos\theta$ is symmetric,
 we assume that the saddle-point  spectral 
density 
$$\rho(\theta) = {1\over N} \sum _{{i=1}}^{N } \delta(\theta - 
\theta_{i})$$
is supported by the symmetric interval $[-a, \ a]$ with $0<a\le \pi$.
The function $\rho(\theta)$ is determined by the saddle point 
equation  
\eqn\sdlptt{
2\lambda  \sin  \theta  =\int_{- a}^{a}
\!  \! \! \! \! \! \! \! \! \! -   d\theta' 
  \rho(\theta')\left(2\cot {\theta-\theta'\over 2} 
 -\cot {\theta -\theta'+i\beta\over 2}
- \cot{\theta -\theta'-i\beta\over 2}
   \right)
}
where we  temporarily rescaled $\b\e \to \b$.
This equation is equivalent to a functional 
equation for the  
 resolvent
\eqn\reZ{ W( \theta) = \hf \int_{- a }^{a} d\theta'  
\rho(\theta')  \cot {\theta - \theta'\over 2},}
 namely
\eqn\cauSat{   \lambda \sin \theta = W(\theta +i0)+ 
W(\theta -i0) -   W( \theta +i \b)
 -   W( \theta- i\b)   }
where $-a<\theta <a  $, supplied with the 
 normalization condition for the density
 \eqn\normden{\oint _{\CC}
€{dz\over 2\pi i}  W(z) = 1}
 where the  contour of integration $\CC$ circles interval
 $[-a, \ a]$.

It is easier to solve this equation for the 
function   
\eqn\zEta{ \eqalign{
\zeta (z)&= -2\cos z + 4 {\sinh{\beta\over 2}\over \lambda}\ \
{W(z +i\b/2) -W(z -i\b/2) \over  i}\cr
&= -2 \cos z
  - {4\sinh^2 {\b\over 2}\over \lambda}  \ \
  \int_{
- a }^{a} {d\theta
\rho(\theta) \over \cos (z - \theta)
-\cosh {\b\over 2}}
\cr}}
which satisfies the simpler equation
\eqn\zETaeq{ \zeta( \theta +i {\b\over 2})= \zeta( \theta -i {\b\over
2})\qquad (\theta\in[-a, a]).}

\def\vy{v_{_\infty}}
The solution can be formulated in terms of standard elliptic functions
(see 
Appendix A for the derivation).
We give it in the form which is convenient for the limit of small $\b$
(or, equivalently, finite $\b$ and $\e\to 0$).
The function $\zeta(z)$ will be given in a parametric form
$$\zeta = \zeta(v), \ \ z=z(v)$$
where the parameter $v$ belongs to the rectangle $- {\pi\over 2 } 
<\r  v<  {\pi\over
2 }, \ -  {\pi\over 2 }\tau < \i v <  {\pi\over 2 }\tau$.
The elliptic modulus $q$ and the nome $k^2$
\eqn\QqQ{ \eqalign{ q &= e^{-\pi K/K'} = e^{i\pi\tau}\cr
k&= \prod_{n=1}^{\infty} \left({1-q^{2n-1}\over 
1+q^{2n-1}}\right)^4
\cr}}
are given below as functions of $\b$ and $\lambda$.

The solution (in parametric form)  is: 
\eqn\ZeTaa{\eqalign{ 
\zeta(v)  \  \ &=    
 { \z _{4} f^2(v)- \z_{3} f^2(\vy)
 \over  f^2(v)  - f^2(\vy) } ,\cr}}
where $f(v)$ is a standard elliptic function
\eqn\snDual{f(v) = {2K' \over \pi}
 \dn\left( {2K' \over \pi} v, \ k'\right)=
1+4\sum_{n=1}^{\infty}
{q^n\over 1+q^{2n}} \cos (2nv),}
and 
\eqn\ZedUU{\eqalign{z(v)  & =  i{\beta\over \pi} v  +
i\ln{\vartheta_1(v+\vy)\over \vartheta_1(v-\vy)}\cr
 & =   i{\beta\over \pi} v + 
i\ln {\sin(v+\vy)\over \sin(v-\vy)}
+4i\sum_{n=1}^{\infty}  {q^{2n}\over 1-q^{2n}}\ 
{\sin 2n\vy \sin 2n v\over n}.\cr
}
}
The  modulus $\tau$ is proportional to the ratio of $\vy$ and $\b$
\eqn\tauvyy{\tau =  4i{\vy\over\beta}, 
\ \ \ \ \ q= e^{- {4\pi\over\beta} \vy},}
 and is determined by  
\eqn\SIXsst{   
   {2\pi\over 
 \lambda}   \sinh ({\beta/2})={\gamma\over 2} 
  E(k) - { \z_{4} \z_{5} +\z_{1} \z_{3} \over 2\gamma} K(k) .  }
The parameters $\z_1, ..., \z_5$ of the solution are expressed
 as functions of  $\lambda$ and $\vy$
as follows:
\eqn\ZeTpp{\eqalign{
  \z_{4}- \z_{3}&=-2 e^{{\beta\over \pi} \vy}{f'(\vy)\over f(\vy)}
   {\theta_1(2\vy)\over\theta_1'(0)} ,\cr
 \z_4+\z_{3}&= -{ e^{{\beta\over \pi} \vy}\over\theta_1'(0)} 
 \left(  { f''(\vy)                                      %
      \over  f'(\vy) } -   {f'(\vy)\over f(\vy)}              %
        +                                                        %
  2  \theta_1'(2\vy)  +{\b\over\pi}\theta_1(2\vy) \right) 
  \cr}}
  \eqn\specpts{\z_{1}= { \z_{3}\alpha^{2} -\z_{4}k^{2 } \over
   \alpha^{2}-k^{2 } }, \ \ \ \z_{5}= {\z_{4}- \z_{3}\alpha^{2}\over
  1-\alpha^{2}}, }
\eqn\qLLpha{\alpha = {f(\vy)\over 1+ 2\sum_{n=1}^{\infty} 
q^{n^{2}€}€},
\ \ \ \gamma = {(\z_{4}- \z_{3} )\alpha\over 
  \sqrt{(1-\alpha^{2})(\alpha^{2}-k^{2})}}}
    Finally, it is useful to know the value $\z_2$ of the function 
$\z(z)$
  at the branch point $z= i{\b\over 2}+a$ 
   \eqn\LamD{  {\z _{4}-\z _{2} \over\gamma} 
={  {\b\over2\pi} +
{\b\over\vy} \sum_{n=1}^{\infty}  
{q^{n}\over 1-q^{2n}} \sin 2n\vy
}.}

\newsec{Scaling limit}

 \subsec{The resolvent in the scaling limit}
  
Let us recall that the parameter $\b$ is the product of the physical
time and the twisting parameter $\e$.  Therefore the twisting is
removed in the limit $\b\to 0$. If $\b\to 0$ with $\lambda $ fixed, we
reproduce the zero-dimensional case considered in \KKN .  In this
section we will consider a  nontrivial limit where both $\lambda$ and
$\b$ go to zero so that the ratio $\lambda/\b$ remains finite.  In
this limit, all observables depend only on the ratio $\lambda/\b$, and
this is why we call it ``the scaling limit''.  Note that in the
thermodynamical limit $N\to
\infty$, the two limits $\lambda\to 0$ and $\e\to 0$ will not commute.

 In the scaling limit we have 
  $\i \tau >>1$,   $K\approx \ln {4\over k'}$, $K' = {\pi\over 
2}(1+{k'^{2}\over 4})$  and, neglecting   the exponentially small 
terms,  we  get
\eqn\solscal{\eqalign{\zeta &= {\z_4+\z_3\over 2}
 -{\z_{43} 
\over 16 q}\ \ 
{1\over \sin(v+\vy) \sin(v-\vy) }
\cr
z &=  i{\beta\over \pi} v + 
i\ln {\sin(v+\vy)\over \sin(v-\vy)}.
 \cr}
}
 where $(\z_{ik} \equiv \z_{i}-\z_{k})$
 \eqn\paranters{\eqalign{
   \z_{43} \  &= 16 q e^{{\b\over \pi}\vy}€  \sin^22\vy \cr
{\z_{4}+ \z_{3}\over 2} &= 
-  e^{{\beta\over \pi} \vy}(2\cos 2\vy    +{\beta\over \pi}  \sin  
2\vy) .\cr}}

 When the regularization is removed,
i.e. in  the limit $\b\to 0$,   
a sensible limit is obtained when $\lambda$ tends to zero 
linearly with $\b$. 
The scaling coupling constant $\b/\lambda$ is obtained from \SIXsst\
after substituting
$E=1, K= {2\pi\over\b}\vy$: 
\eqn\lambdavy{2\pi{\b\over \lambda}   
= 2\sin 2\vy
 - 4\vy \cos 2\vy  .
}

  \subsec{The free energy in the scaling limit}

The derivative of the free energy 
$$F(\lambda, \beta) = \lim {1\over N^2} \ln \CZ_N(\lambda, \beta)$$
is 
proportional to the first moment of the spectral density
\eqn\Fprim{\eqalign{
F'_{\lambda}(\lambda, \beta) &=   
\int_{-a}^{a}d\theta 
\rho(\theta) \cos
\theta ,\cr }}
which can be evaluated by looking at expansion 
of $\zeta(z= \pi +iy)$, at $y\to\infty$,
$$\zeta (z) = e^{y}+ \sum _{k=1}^{\infty}   \zeta^{{(k)}}\ e^{ 
-(2k+1)y} .$$
We have 
\eqn\Fprimm{\eqalign{
{F'_{\lambda}(\lambda, \beta) }&=  {\lambda \over 8 
\sinh^2{\b \over 2}}\ \ 
[\z^{(1)}-1].\cr}}
The coefficient $\zeta^{(1)}$ is evaluated 
 in the scaling limit $\b\to 0$  in Appendix A.
 This allows us to write  an explicit   expression 
for the free energy in the  limit $\b\to 0$:
\eqn\frenSL{{F'_{\lambda}(\lambda/\beta)   }
=   {\lambda \over 4\pi \b}( 4\vy -\sin 4\vy), \quad
  \pi{\b\over \lambda}   
= \sin 2\vy
 - 2\vy \cos 2\vy  .
}
This expression for the free energy is universal in a certain sense:
if one deforms the potential \poto\ to a more general one:
\eqn\potn{  \la \Tr \sum_{n= -\infty }^\infty t_n \Omega^{n} ,
} 
then the scaling limit of the free energy   will
have the same form \frenSL, where $\la$ will be substituted by some 
function of the 
couplings
$\tilde\lambda(g_1,g_2,\cdots)$.  
The universal form of the free energy can only change if we tune
  the couplings $g_n$ to some multicritical point.

The corrections to the eq. \frenSL\  are of two kinds:  power-like
corrections  and  exponentially small terms of the type $$q
= e^{-4\pi \vy/\b}.$$ In the limit $\lambda \ra \infty$ we have
$q=\exp -\left({24 \pi^4\over \lambda\be^2}\right)^{1/3}$. These 
terms 
are of
course invisible   compared with the power-like corrections but
they imply the existence of essential singularity in the $\b \ra 0$.

If we return to the original notations   in terms of $\beta$,$\e$ and $\la$
we conclude from (5.6) that $F(\la,\be,\e)=\eps\lambda f(\la/(\be\eps))$.
Hence the principal $\sim N^2$ correction to the free energy tends to zero
in the limit $\eps\ra 0$ (when we recover the original unperturbed reduced
SYM theory). On the other hand, as we will see below from (5.6), there is
no regular expansion in powers of $\e$ in the weak coupling phase which
signifies that there is an essential singularity at the origin of this
coupling and, correspondingly, in the moduli space of our theory.

\subsec{ The Gross-Witten phase transition and the strong coupling 
phase}

The matrix integral we are considering has qualitatively the same
phase structure as the   $U(\infty)$ gauge theory on a 
two-dimensional sphere.  The weak coupling
phase considered above, describes the range of couplings $\lambda >
\lambda_{c}$ where
\eqn\lavycr{\lambda_{c}= {1-e^{-\beta }\over \e}}
is determined by from the condition $a=\pi$, 
i.e. that the two endpoints of 
the cut meet on the unit circle. 
(Eq. \lavycr \ follows from \lambdavy \ with
$\vy=\pi/2$; the length of the cut as a function of 
$\lambda$ is given in Appendix A.)
The  singularity  near this critical point is as usual of third order.

 The strong coupling solution is obtained by expanding the 
 spectral density and the kernel in a Fourier series
 $$\rho(\theta) ={1\over 2\pi} \sum_{n=0}^{\infty}c_{k}\cos 
(k\theta).$$
 $$  \cot {\theta + i0 \over 2 } +
 \cot {\theta  - i 0\over 2}  - \cot {\theta + i\beta \over 2 }
 -\cot {\theta - i\beta \over 2 }= 
 4  \sum_{k=1}^{\infty} [1 - e^{-k\beta}]  
  \sin(k\theta) 
 $$
 $$
 \int_{-\pi}^{\pi}
\!  \! \! \! \! \! \! \! \! \! -  \  d\theta' 
  \rho(\theta') \cot {\theta-\theta'\over 2} 
  =   \sum_{k=1}^{\infty}  c_{k}
  \sin(k\theta) $$
  
   It is therefore clear that only the  $c_{0}$ term  of the 
   expansion of the spectral density has to be retained.
   One finds
   \eqn\spdSCF{ \rho(\theta) = {1\over 2\pi} \left( 1 + {\lambda 
   \over 1- e^{-\beta}} \cos \theta \right) \qquad {\rm for}
   \qquad 0<\lambda < \lambda_{c}={1-e^{-\beta } }.}
  
For the free energy we find then:
\eqn\Fsc{ {F'_\lambda }= \int_{-\pi}^{\pi}  d\theta 
  \rho(\theta) \cos \theta = {1\over 2}\  {\lambda\over 1-e^{-\beta}}
\qquad {\rm for} \qquad 0<\lambda < \lambda_{c} }
In the scaling limit $\e\to 0$  we obtain:
$  F'_\lambda  = {1\over 2}   {\lambda\over\beta}$ \ for 
$\lambda < \beta$.
At the critical point  
$\lambda_c = \beta$ we have $F'_\la|_{\la=\be} ={1\over 2}   $ 
and 
$  F''_{\la }|_{\la=\be}={1\over 2\beta} $.
 A simple calculation  using  eq. \frenSL\ gives in the weak coupling
 phase
 $\lambda > \lambda_c$
 the same values of first two derivatives 
of the free energy at the critical point. This means that we have, as
usually, the 3-rd order Gross-Witten phase transition. Note that in
the limit $\beta \ra \infty$ our model reduces indeed to the
 one-plaquette
 model originally studied by Gross and Witten \GW.

   
\subsec{ Reduction to the zero-dimensional theory:  $\beta <<\lambda$}

In this limit the  theory appears to be the zero-dimensional
reduction of $CN=1$ SYM studied in \KKN.  The integral \twpa\ reduces
(after the rescaling $\theta
\ra \beta\theta$) to  a simpler integral:
\eqn\twpc{ \CZ_N(\beta,\lambda)= {2\pi N\over N!} {\be\over 2\pi}^N 
e^{N\lambda}
\int \prod_{k=1}^N {d\theta_k \over 2\pi} \   e^{-\hf N\lambda
\beta^2 \theta_k^2}\ \prod_{i\ne j} { ( \theta_i - \theta_j)
\over  ( \theta_i - \theta_j +i )}}
  
This model was studied in \Hoppe\ and later in \KKN .

We find from \frenSL\ the following
expansions in half-length of the cut $a=2\vy$:
\eqn\Fexps{ F'_\lambda = 1-{a^{2} \over 10}   + {a^4\over 
4200}+\CO(a^{6} )}   
\eqn\xiexps{   {\b\over \lambda } = {a^{3}\over 3\pi}(1- {a^{2} \over 
10}) +  
\CO(a^{7})
  }
This gives the following asymptotics for the free energy:

\eqn\asy{ F(\b, \la) =
\la \left[ 1-  {3\over 10} \left( {3\pi\beta\over \lambda} 
\right)^{2/3}
-{27\over 1400}\left( {3\pi\beta\over \lambda} \right)^{4/3} 
 +O \left( {\beta^2\over \lambda^2}\right)\right]}
The first term of this expansion matches   the asymptotics of big 
$\lambda$ obtained in \KKN\ for the integral
\twpc\ in the large $N$ limit. The next terms are not supposed to 
match with \KKN\  since we already used the scaling limit expression 
of
the free energy with the finite compactification radius $\beta$.
 

\newsec{ Conclusions}

Let us outline the main results of the paper:

\noindent
1. We consider a topological sector of the one-dimensionally reduced
$\CN=1$ $SYM_4$ theory on the light-cone time circle with a  a special
massive
perturbation. Using   Wittens nonabelian  localisation principle
\witloc,  we represented the partition function with periodic b.c.
 in terms of a
solvable matrix quantum mechanics (twisted matrix oscillator).

\noindent
2. We find the integrability properties of this model relating it to
the Toda hierarchy. The generating functional as a function of its
parameters satisfies the Toda chain equation.

\noindent
3. In the large $N$ limit we find the exact solution of the model: the
generating functional is parametrized in terms of elliptic functions.
We find the Gross-Witten type phase transition and identify its
location. The strong coupling solution is also found.

\noindent
  4. An interesting model  corresponds to the analytical continuation
$\beta \ra i\beta$ (inverted matrix oscillator) being known to have
the properties of the $c=1$ non-critical strings.

\noindent
5. In the scaling limit of vanishing perturbation we find a simple
universal (with respect to certain deformations of parameters of the
generating functional) expression for the free energy and Wilson loop
correlators along the light-cone circle. Its strong compactification
limit restores similar results for the completely
reduced $\CN=1$ $SYM_4$ considered in \KKN .

Some remaining problems:

\noindent
1. We need further understanding of the space time symmetries of the
model and of the  correlators corresponding to our generating
functional. 

\noindent
2. The representations similar to the eq.\twpa\ for the $SYM_4$ can be
found also for the $SYM_6$ and $SYM_{10}$ reduced to the light-cone 
time
circle: we just have to take the corresponding eigenvalue integrals
for the partition functions in the paper \KKN and substitute there the
rational functions by trigonometric  ones.   Unfortunately, we
cannot apply the powerful methods used here to those models: we don't
know any relation of them to the integrable hierarchies and we cannot
solve exactly the large $N$  saddle point equation. On the other 
hand, it
seems to be possible to investigate this saddle point equation in the
scaling limit similar to tha used in the present paper.

Another interesting question is whether two prescriptions for the 
contour integration with respect to the eigenvalues of $\phi$ 
coincide for
the infinite $N$  in some part of the phase space of parameters 
$\e\beta$
and $\lambda$. It is clear that they give different answers in the
strong coupling phase $\lambda < \lambda_c(\beta)$ (since they are
already different for $\lambda=0$). As for the weak coupling phase
$\lambda > \lambda_c(\beta)$ of our large $N$ solution (which is not
even analytical at $\lambda=0$),  it is possible that the saddle point
approximation does not distinguish between two different prescriptions
of integration over $\phi(t)$ (contour integration over the Cartan
subalgebra, on the one hand, and integration over $\theta$'s in the
finite interval $[0,2\pi]$, on the other hand). This hypothesis is
  to be verified. A weaker version of it could be the coincidence of 
two
prescriptions in the scaling limit (eq. \frenSL ).

\newsec{Acknowledgments}
%
 We are  grateful to 
N. Nekrasov for many valuable advices and in particular for 
explaining to  us the 
construction  of refs.
\refs{\NEKdiss ,\NekBau}. We also thank A.S.Schwarz, S.Shatashvili 
and 
A.Wainstein for useful discussions.
One of us (J.H.) would like to thank J. L. Gervais for his kind 
invitation to LPT ENS in January 1999.
 This research
is supported in part by European  TMR contract ERBFMRXCT960012.


\appendix{A}{Solution of the saddle point equations}

%
\def\sn{{\rm sn} }

\def\dn{{\rm dn} }

\subsec{The function  $z=z(\zeta)$ 
as an elliptic integral}
 
 It follows from the integral 
  representation of $\zeta(z)$ that
 it
is real  when 
$z \in \IR,  i\IR,  i \IR \pm  \pi $, 
satisfies 
\eqn\propaa{ \zeta(z) =\zeta(z+2\pi)=
\zeta(-z) = {\overline{\zeta({\zb})}}  }
and by 
\zETaeq\  is also real along the interval
$[ {i\over{2}}\b  - a,
  {i\over{2}}\b  +a]$.
Therefore this function defines a map of the 
half strip $0< \r z < \pi, \i z >0 $
with a cut  $[ {i\over{2}}\b ,
  {i\over{2}}\b  +a]$, to the upper half plane $\i \zeta >0$ (Fig. 1).
  The inverse map $z=z(\zeta )$ is given by
the Schwarz-Christoffel  formula
(see, e.g. \FS  ): 
\eqn\mapp{z=i\int_{\z_{4}}^{\zeta }
{dt\  (t-\z_{2})\over Y(t) }    }
where
\eqn\Ry{Y(t) =  
 \sqrt{(t- \z_{1})(t-\z_{3})(t-\z_{4})(t-\z_5) }\ \ .}
By construction, the map \mapp\  acts on  the special
  points $\z_{1} < \z_{2 } < \z_{3} < \z_{4}<\z_5$
 and $\infty$
as is shown in the two first coloumns of Table 1.

The  values of $\zeta$ at the special points of the map 
are determined as functions of $\b$ and $\lambda$ by the assymptotics 
of $\z(z)$ at infinity.
The expansion of the  function \zEta\
at $z\to\infty$ contains only odd 
powers of $e^{iz}$. 
If we approach infinity  as 
$z=\pi + i y,\  y\to\infty,$  the asymptotics of 
$\zeta (z)$ is 
\eqn\sedem{
\zeta_{+}(y)\equiv 
 \zeta (\pi +iy)=e^{y} + \sum_{n=0}^{\infty} \zeta ^{(2n+1)} 
e^{-(2n+1)y}}
 where
 \eqn\sedembis{\zeta ^{(1)} = \left(1+{8\sinh^{2}{\b\over 2} \over 
\lambda}
\int_{-a}^{a}d\theta \rho(\theta) \cos \theta \right), }
etc.

\subsec{Elliptic parametrization of the 
 solution}

The map \mapp\  and the condition \normden\ 
can be expressed 
explicitly
in terms of 
standard elliptic integrals  (see \BF, 256.02 ) with  parameters
  
     \eqn\kakaprim{
k   = \sqrt{\z_{54}\z_{31} \over  
\z_{53}\z_{41}} , \ \  k' = \sqrt{1-k^{2}} = \sqrt{\z_{43}\z_{51} 
\over  \z_{53}\z_{41} }}
\eqn\gammanu{\gamma =  \sqrt{\z_{53}\z_{41}},\  \
\alpha^2 =  {\z_{54}\over \z_{53}}, \ \ \ \nu =
   \arcsin \sqrt{\z_{41}\over \z_{51}}}
 where the notation $\z_{ij}= \z_{i }-\z_{j}$ is used.  Namely
 \eqn\ZetaZ{\eqalign{z(\zeta) &= {2\z_{43}\over \gamma}\left( {\sl 
\Pi }(\varphi, 
 \alpha^{2},
 k) +  {\z_{32} \over \z_{43} }F( \varphi, k)\right)\cr
   \varphi & = \arcsin {1\over \alpha}
\sqrt{  \zeta - \z_4\over \zeta - \z_3}, \cr}}
and
\eqn\SIXsst{   
   {2\pi\over 
 \lambda}   \sinh ({\beta/2})={\gamma\over 2} 
  E(k) - { \z_{4} \z_{5} +\z_{1} \z_{3} \over 2\gamma} K(k).   }
 It is convenient to introduce as a parameter the elliptic amplitude 
 $u$
 related to  the angle $\varphi$ as
 $ \sn u = \sin\varphi $ 
\eqn\oMegA{   u  \  = 
i{\sqrt{\z_{53}\z_{41}} \over 2} \int_{\z_{4}}^{\zeta }
{dt  \over Y(t) } =  
F(\varphi, k) , \ \ \ \sn u 
=\sin \varphi
={1\over \alpha}
\sqrt{  \zeta - \z_4\over \zeta - \z_3} .   }
Then the function 
\eqn\ZeTaa{ \zeta(u) =    
 { \z_{4} - \z_{3} \alpha^2 \sn^2 u
 \over  1 -   \alpha^2 \sn^2 u
 } }
maps the upper $\zeta$-half-plane is mapped    to 
the rectangle  $0\le \r u \le K, \ 0\le \i u \le K'$
with  $K$ and $K'$ being the complete elliptic integrals
associated with the moduli $k$ and $k'$.
 The  special points  $\zeta = \z_{1}, \ldots, \z_{5} $  and $\infty$
 correspond to the points
 $u_{1}, \ldots, u_{5}$ and $u_{\infty}$  along the boundary of the 
rectangle   
as is shown in Table 1.   
 Note that
$${1\over\alpha} = \sn(u_\infty).$$
 \vskip 20pt
 \hbox{\qquad\qquad\qquad\qquad\qquad\qquad
\vbox{\offinterlineskip
\hrule
\halign{&\vrule#&\strut\quad\hfil#\quad\cr 
height2pt&\omit&&\omit&&\omit
&&\omit&\cr
& $\zeta $ \hfil && $z$ \hfil && $u$   \hfil && $v$ &
 \cr
height2pt &\omit &&\omit&&\omit&&
\omit&\cr
\noalign{\hrule}
height2pt&\omit&&\omit && \omit&& \omit&\cr
& $-\infty$ && $+i\infty$ && $u_{\infty} +i0
$&& $ v_{\infty} +0 $
&\cr
& $\z_{1}$ && $ {i\over{2}}\b$  && $ K+iK' $&& ${\pi\over 2} $
 &\cr
& $\z_{2}$&&${i\over{2}}\b +a$&&$u_{2} $&&$v_2
$&\cr
& $\z_{3}$ && ${i\over{2}}\b  $  && $  iK' $&& ${\pi\over 2}(1+\tau)$
&\cr
& $\z_{4}$ && $0$  && $0 $&& ${\pi\over 2} \tau $
&\cr
& $\z_5$ &&$   \pi    $  && $K  $&& $0  $
&\cr
& $+\infty$ &&$ \pi  +i\infty$  && $u_{\infty}+i0
$&& $ v_{\infty} -0 $
&\cr
height2pt&\omit&&\omit&&\omit&&\omit&\cr}\hrule}
}

\vskip 11pt
 
\centerline{Table 1. The values of 
$z, \zeta, u$ and $v$ at the special points of the map.}

\vskip20pt
 
The function  $z(u)$ 
 defined by the integral \mapp \ 
  reads, in the  parametrization \oMegA ,
\eqn\mappp{ z(u)      = {2\z_{43} \over \sqrt{\z_{53}\z_{41}}}
\left( \int_0^u {du\over 1-\alpha^2 \sn^2u} + {\z_{32}\over \z_{43}} u
\right).
}
  We will express the integral in  \mappp\  in terms 
of Jacobian elliptic functions. 
   Since 
the point $ u_\infty $ is   between $u_{5}$ and   $u_{1}$),
it has the form
\eqn\Uxi{ u_\infty = K+i\xi, \ \ 0<\xi  < K'.}
  We find,     using eq. 433.01 of \BF ,
 \eqn\ZzZU{z(u)= i\ln {H_1 (  u+i\xi   )
\over H_1(  u  - i\xi)}
+  {u\over K}  \left[
{2\z_{42} \over \sqrt{\z_{53}\z_{41}}} K 
-\pi  \Lambda_0(\nu, k)\right]
}
 where 
$ H_1(u) $
is   a standard Jacobian elliptic function and 
$$\eqalign{ \Lambda_0(\nu, k)&= {2\over \pi}
[(E-K)F(\nu, k') +KE(\nu, k')]
\cr &= i{H_1'(i\xi)\over H_1(i\xi)}, 
\ \  \ \ \left(\nu = 
   \arcsin \sqrt{\z_{41}€\over \z_{51}}
=\arcsin  { \sqrt{\alpha^2-k^2}\over k'}\right)\cr}
$$
is known as the Heuman's Lambda function.
The condition   $ z(iK') = z(K+iK')$
is satisfied only if the coefficient in front of the
linear term in $u$ is zero, hence the condition 
\eqn\lAMB{ {2\z_{42} \over \sqrt{\z_{53}\z_{41}}} K 
=\pi  \Lambda_0(\nu, k)}
 From ${H_1(u+iK')  }= e^{-i\pi u\over K} H_1(u-iK' )$
 we find  
 $$ z(i K') = i {\pi \over K} \xi $$
which
 allows to determine  $\xi$:
\eqn\xiBETA{      \xi  = {
K\over2\pi}   \beta .}
  The final expression for $z(u)$ is  therefore
\eqn\ZotU{ z(u)= i\ln {H_1 (  u+i {K\over 2\pi} \beta  )
\over H_1(  u  - i {K\over 2\pi}\beta)}.}

 \subsec{ The dual modulus}

We are going to write our solution in  a form, which  
will allow to perform  painlessly the scaling  limit
${\b } \to 0$.  In this limit 
$a_{43}\approx -i \b {d\zeta\over dz}\to 0$ and,
according to \kakaprim , $k' \approx 4 e^{-K} \to 0$.
Therefore it is more convenient to  expand 
 the solution in the dual modular parameter
\eqn\QqQ{ \eqalign{ q &= e^{-\pi K/K'} = e^{i\pi\tau}\cr}}

and use the   variable $v$
$$ v = {\pi\over 2} \tau -i { \pi/2 \over  K'}\ u
$$
as a parameter.
The elliptic nome is expressed as a function of $q$ as
\eqn\NOme{
\eqalign{k&= \prod_{n=1}^{\infty} \left({1-q^{2n-1}\over 
1+q^{2n-1}}\right)^4
\cr}}
 The parameters corresponding to the special points of the map are 
given by the last coulumn of Table 1.
The parameter of infinity is equal, by \Uxi\ and \xiBETA , to 
\def\vy{v_{_\infty}}
\eqn\vyB{\vy   
= {K\over 4K'} \b = 
-{i\tau\over 4} \b
\ \ \ \ \ (0<\vy<{\pi\over 2}).}
The domains of the four variables $z, \zeta, u$ and $v$ are depicted 
in Fig. 1.

We will write the solution as
a function of the parameters
 $\vy= \vy(\lambda)$  and $\beta$.  It will be written  as a series 
in the expansion parameter $q$
\eqn\tauvyy{ 
 q= e^{- {4\pi\over\beta} \vy} = e^{i\pi \tau}, \ \ \ \ \ \tau =  
4i{\vy\over\beta}.}
The expansion of the function $\zeta(u)$ is 
obtained by plugging in
eq. \ZeTaa\
the  representation of $\sn$ in terms of the dual modulus
\eqn\snDual{{1\over \sn u} = 
 \dn\left( {2K' \over \pi} v, \ k'\right)={\pi\over 2K'}f(q)}
\eqn\fqU{f(q)= \left[1+4\sum_{n=1}^{\infty}
{q^n\over 1+q^{2n}} \cos (2nv)\right].}
  The function $z(v)$ reads, in terms of the standard elliptic 
functions associated with the dual modulus, 
\eqn\ZedUU{\eqalign{
z(v)  & = -{4 \over \pi\tau} \vy v +
i\ln{\vartheta_1(v+\vy)\over \vartheta_1(v-\vy)}\cr
 & =   i{\beta\over \pi} v + 
i\ln {\sin(v+\vy)\over \sin(v-\vy)}
+4i\sum_{n=1}^{\infty}  {q^{2n}\over 1-q^{2n}}\ 
{\sin 2n\vy \sin 2n v\over n}\ .\cr
}
}
   
 Finally, \lAMB\ expands as
\eqn\LamD{{4\vy \over \b} {\z_{42} \over \sqrt{\z_{53}\z_{41}}}
 ={2\over \pi} \vy+  {\vartheta_4'( \vy)\over \vartheta_4( \vy)}
={2\over \pi} \vy +
 4\sum_{n=1}^{\infty}  
{q^{n}\over 1-q^{2n}} \sin 2n\vy .
}

In order to fix completely the solution, let us consider the vicinity 
of the point $\vy$
and compare the explicit dependence $\zeta = \zeta(z)$
with the asymptotics  \sedem
at $z\to  \pi+ i\infty$.
The half-line  
 $$ z = \pi + iy   \ \ \ (y>0)$$
is parametrized by the interval $0<v<\vy$.
 In the left vicinity of the point $\vy$
$$v =\vy -\epsilon    \ \ \ (\e >0)$$ 
the functions $z = \pi +iy $ and $\zeta$ have the form 
  $$
  e^{y } = {A\over \e} +B +C\e, \ \ \  \zeta = {P\over \e} +Q+R\e$$
 with 
$$ P= -\z_{43} {f(\vy)\over 2 f'(\vy)},
\ \ Q= - P\left( 2{\z_4\over \z_{43}} { f'(\vy)\over
f(\vy)}-\hf   { f'(\vy)\over f(\vy)}
- \hf   { f''(\vy)\over f'(\vy)}\right),$$
 $$A =  e^{{\beta\over \pi} \vy} {\theta_1(2\vy)\over\theta_1'(0)} , 
\ \ 
  B = - A\left({\b\over \pi} + 
{\theta_1'(2\vy)\over\theta_1(2\vy)}\right).$$
 
 The  leading 
 asymptotics \sedem\ of $\zeta(z) $ is achieved if
  $ A=P$ and  $B = Q$, which yields \ZeTpp .

>From \LamD\  we get
\eqn\achdv{\z_{42}  =\sqrt{\z_{53}\z_{41}}
{\b \over 2\pi} \left( 1+   2\pi   q { \sin 2\vy \over 
\vy}\right) }
  (Note that the relation $2\z_{2} = \z_{1}+\z_{3}+\z_{4}+\z_{5}$ is 
satisfied.)

 %
\vskip 50pt
\hskip 20pt
\epsfbox{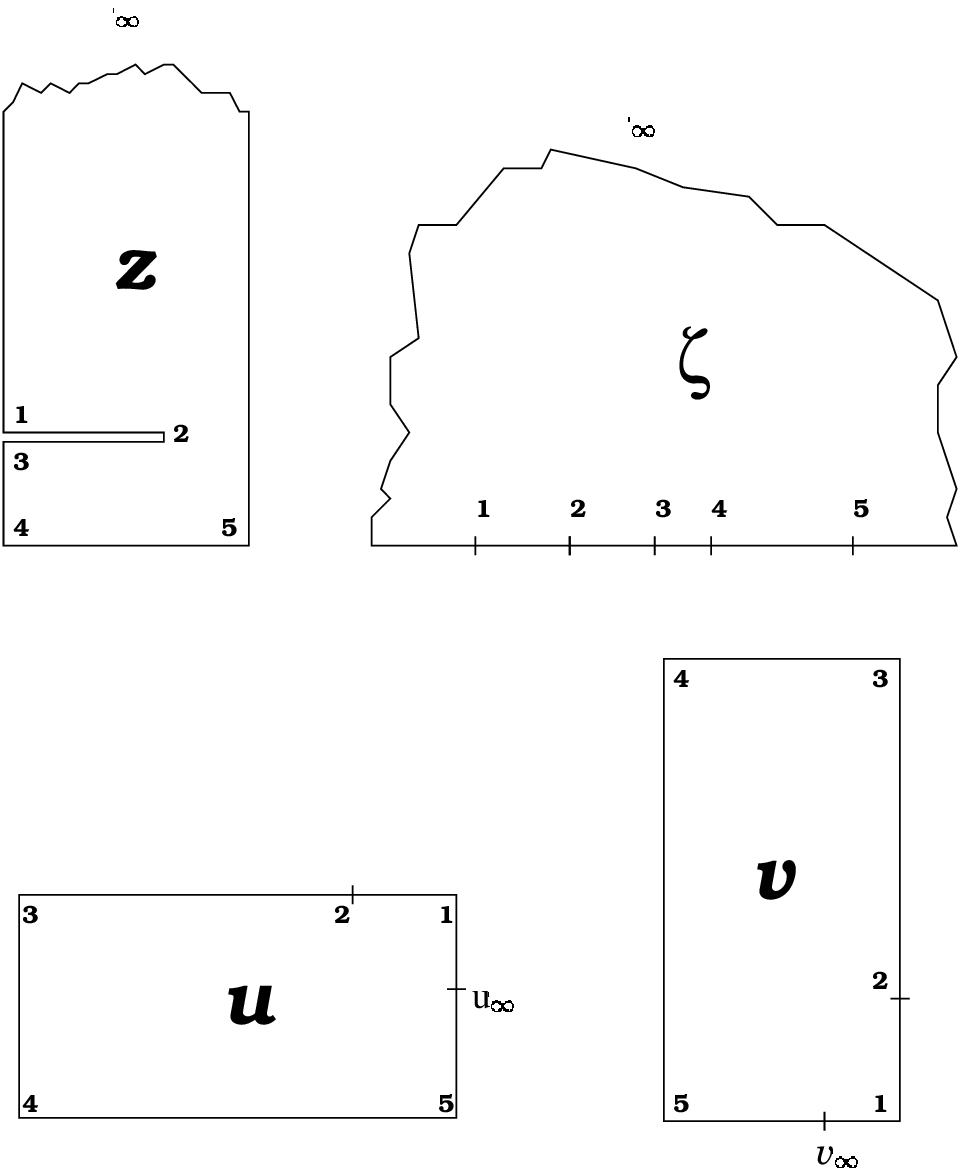}
\vskip 20pt
\centerline{Fig.1:  The domains of the variables $z, \zeta, u, v$.  }

\bigskip

\subsec{The limit of large $\tau $ (small $\b$)}

In this limit, which corresponds to the scaling limit
discussed in Section 4, the $v$-rectangle can be replaced by an 
infinite half-strip and elliptic functions degenerate to 
trigonometric functions.
 After substituting
$E=1, K= {2\pi\over\b}\vy$ in  the normalization condition \SIXsst , 
 we get in this  limit
 \eqn\lambdavy{{4\pi\over \lambda} \sinh {\beta\over 2} 
e^{-{\beta\over\pi} \vy}
= 2\sin 2\vy
 - 4\vy [\cos 2\vy - {\beta\over 2 \pi} \sin 2\vy].
}
The parameters of the solution are obtained from
$$P={\z_{{43}}\over 16q}\ {1\over \sin 2\vy}, 
  \ \ \ 
  Q = \z_{4}+      P \cot 2\vy, \ \ \ 
  R = P\left( {2\over 3} + \cot^{2} 2\vy\right)$$
$$A =  e^{{\beta\over \pi} \vy} \sin 2\vy , \ \ \ 
  B = - A
\left( \cot 2\vy  +{\beta\over \pi}  \right), 
  \ \ C = A\left( - {1\over 3} + {\b\over \pi}
\cot 2\vy + {\beta^{2}\over 2\pi^{2}}\right).
  $$
 From $A=P$  we get
  \eqn\achtri{ \z_{43}= 16 q e^{{\beta\over \pi} \vy}\sin^{2} 2\vy
 }
  and   $B=Q$    implies
  \eqn\achtr{
 {\z_3+ \z_{4} \over 2} 
= -  e^{{\beta\over \pi} \vy}(2\cos 2\vy    +{\beta\over \pi}  \sin  
2\vy)
= -\left( e^{{\beta\over \pi} \vy} \sin  2\vy\right)'_{\vy}
.}
 It is useful to note that
$${\z_{53}+ \z_{54}\over \z_{41}+ \z_{\z1}}= \cot^{2}\vy, \ \ \ 
{\z_{42}+ \z_{32}\over \z_{41}+ \z_{31}} = {\b\over 2\pi} \cot\vy ,
\ \ \ {\z_{43}\over \z_{41}+ \z_{31}} = 16 q 
\cos^{2} \vy.$$
Finally, the coefficient $\z^{(1)}$ is obtained as
\eqn\Fprim{\eqalign{\z^{(1)}&=
A(R-C)\cr
 =& 1+  
   2{\b}^{2}      
 \left( {4\vy - \sin 4\vy \over 4 \pi \beta\e }  +  { 4\vy^{2}
 - \sin^{2} 2\vy - 2\vy \sin 4\vy\over 8\pi^{2}} + 
\CO(\b)\right).\cr}}

\subsec{The length of the cut}

The branch point of the Riemann surface of $\z(z)$ is at $ z_{2} = 
z(v_{2}) $, where $ \z (v_{2})=\z_{2}$.
   Taking the limit of \achdv, 
\eqn\achdve{ 
\z_{42}  =
  {\b \over  \pi} \sin 2\vy e^{{\beta\over \pi} \vy}
}
 we rewrite  the solution  \solscal\ in the 
form

\eqn\solexp{\eqalign{                                                 
z&=  i{\beta\over\pi}v + i\ln {\cot   v +\cot 
\vy\over                   \cot   v -\cot \vy}\cr                                                                                \cot^{2}v 
& = { (\z_{4} -\zeta ) \cot \vy +{2\pi \over\b}
\z_{42}\over                                          (\z_{4} -\zeta 
) \tan \vy -{2\pi \over\b}  
\z_{42}}                                                     
.\cr}}                           
 Putting  $v=v_{2}$ in \solexp, we get  
   $$a=-{\b\over\pi} \delta_{2} + i\ln { \tanh  \delta_{2} -i\cot 
\vy\over 
 \tanh  \delta_{2}  +i\cot \vy},$$
   $$\tanh^{2}\delta_{2}€ =  { 1- {\beta\over 2\pi} \cot \vy
 \over 1+{\beta\over 2\pi} \tan \vy}$$
and finally
 $$\cos(a+{\b\over\pi}\delta_{2})= \cos 2\vy - {\b\over2\pi}\sin 2\vy
 \approx  \cos 2\vy , $$
which allows us to evaluate $a$
\eqn\cutaa{a\approx 2\vy - {\b\over\pi}\delta_2, \ \ \ 
\delta_2\approx \ln\left( {4\pi\over\beta} \sin 
 2\vy\right). }

\appendix{B}{Direct scaling analysis of the equations on parameters 
of 
the large $N$ solution }

Six conditions on the length of the cut, $a$, and the 5 parameters of
the map $z\mapsto\zeta(z)$ (which we denote here by $a_{1}, \ldots,
a_{n}$ instead of $\z_{1}, \ldots,
\z_{n}$)
are
\eqn\sixconds{\eqalign{
(1) \qquad & a  = 
   \int_{a_{2} }^{a_{3} }  dt  {
t-a_{2} 
\over
|Y(t)| } \cr 
 (2) \qquad  & 0  =  
  \int_{a_{1} }^{a_{3} } dt 
{  t-a_2\over |Y(t)|} \ \ \ \ \left({\rm resp.} \ \ \ \ 
\pi   =    \int_{a_{4}}^{a_{5}} dt {t -
a_{2}\over|Y(t)|} \right)\cr
(3) \qquad & \half\b  =    \int_{a_{3} }^{a_{4}} dt {t- 
a_{2}\over|Y(t)|}\cr 
(4) \qquad & a_{1} +a_{3} +a_{4}+a_{5}  = 2a_{2}  \cr
(5) \qquad &\ln a_{5} = \int_{a_{5}}^{\infty} 
 dt \left(
{t- a_{2} \over{Y(t) }} -{1\over t}\right)
 \ \ \ \ ( a_5>0),\cr
(6) \qquad &{4\pi \over\lambda} \sinh  (\beta / 2)
  =\int_{a_{4}€}^{a_{5}}d\zeta {
  \zeta (\zeta-a_{2})\over  |Y(\zeta)
| }\cr}}
with $Y(t)$ as in (A3) (where the $a_i$ were denoted by
$\zeta_i$). While the first three conditions are implied by the
geometry of the map $\zeta$, conditions (4) and (5) follow when
comparing the (from (4.5)) known asymptotics of $\zeta$, e.g. for
$z=\pi+iy$ ($y\to\infty$; cp. (A4)), to the one implied by the
integral representation (A2) which says that
\eqn\dingens{y=\ln\zeta_+ +\left(\int_{a_5}^\infty dt\left( 
{t-a_2\over Y(t)}
-{1\over t}\right)-\ln a_5\right) -{1\over\zeta_+}\left(
{1\over 2}(a_1+a_3+a_4+a_5)-a_2\right) -\sum_{n=2}^\infty
{b_n\over\zeta_+^n}}
with $b_2 ={1\over 16}(a_1^2+a_3^2+a_4^2+a_5^2-\prod_{2\ne i\ne j\ne 2}
a_ia_j)$ already determining 
$\int_{-a}^{+a}d\theta\,\rho(\theta)\cos(\theta)$.

With
  \eqn\gammanu{a_{ij}=a_i-a_j,\ \ \gamma =  \sqrt{a_{53}a_{41}},\  \
 \ \ \ \nu = 
   \arcsin \sqrt{a_{41}€\over a_{51}}
   k   = {\sqrt{a_{54}a_{31}}\over \gamma}, 
\ \  k' = \sqrt{1-k^{2}} = {\sqrt{a_{43}a_{51}}\over \gamma},}
the   conditions (2), (3), (5) and (6)    read %
\eqn\SIXdve{\eqalign{
 (2) \qquad K(k) &  = {a_{43}\over a_{42}}
{ \sl \Pi} \left( {a_{31} \over a_{41}}, \ 
k\right)\cr
   &= {a_{51}\over a_{52}}{\sl \Pi} \left( -{a_{31}\over a_{53}}, \ 
k\right) 
  \cr
    &= - {a_{42}\over a_{32}}
{\sl \Pi} \left( {a_{54}\over a_{53}}, \ k\right) +
     {  \pi\gamma \over 2 a_{32}} 
  \cr
}}
\eqn\SIXctri{\eqalign{ (3) \qquad 
\hf \beta  &=
{2a_{31}\over  \gamma} { \sl \Pi} \left( {a_{43} \over a_{41}}, k' 
\right) -
{ 2a_{21}\over  \gamma} K(k')\cr
&= - {2a_{54}\over  \gamma}  { \sl \Pi} \left( {a_{43} \over a_{53}}, 
k' \right)
  +{2a_{52}\over  \gamma} K(k') \cr
}}
\eqn\SIXpet{\eqalign{ (5) \qquad 
 {2a_{54}\over \gamma} 
 \sl \Pi} \left( \nu
,\ {a_{43} \over a_{53}},\ k' \right)
- {2a_{52}\over \gamma} 
 F(  \nu, k')=
  \ln \left( {a_{53}+a_{41}  \over 
4}\right)}
\eqn\SIXsst{  (6) \qquad 
   {2\pi\over 
 \lambda}   \sinh ({\beta/2})={\gamma\over 2} 
  E(k) - { a_{4} a_{5} +a_{1} a_{3} \over 2\gamma} K(k).}

The scaling limit can be obtained by direct analysis 
 of the equations  \sixconds\ 
 in the limit
  $\beta \to 0, 
a \to a_{0}>0$.  
Letting
$$
a_{43} = u, a_{32} = v, a_{53} = w , 
$$
$$
 \tilde u = {u\over a_{31}},
\tilde v = {v\over a_{31}},
\tilde w = {w\over a_{31}},$$
one has
\eqn\Sctri{ (3) \qquad 
\beta =  2
\int_0^1 ds { \tilde v+ \tilde u s\over
\sqrt{ s(1-s)(\tilde w -\tilde u s)(1+ \tilde u s)}}}
$$= {2v\over wa_{31}}\int_0^1 ds {1+{u\over v}  s\over
\sqrt{ s(1-s)(1-{u\over w}  s)(1+ {u\over a_{31}} s)}}$$
\eqn\SeONE{ (1) \qquad
a = \tilde v  
\int_0^1 ds { 1-   s\over
\sqrt{ s(s+{\tilde u\over \tilde v})(\tilde w +\tilde v s)(1- \tilde 
v s)}}}
$$ = {v\over wa_{31}}\int_0^1 ds {1-s
\over
\sqrt{ s(s+{u\over v})(1+{v\over w}  s)(1- {v\over a_{31}} s)}}
$$

 \eqn\Sdve{ (2)\eqalign{ \qquad    
 \pi &= {v\over \sqrt{w a_{31}}} 
 \int_{{u/w}}^{1} ds {1+{w\over v} s\over
 \sqrt{ s(1-s)(s-u/w) (1+s\tilde w)}} \cr
&\approx {2\tilde v\over \sqrt{\tilde w}}K(r) +
\sqrt{w} \int_0^1{ds\over \sqrt{(1-s)(1+s\tilde w)}}\cr}
}
where $r^2 =(1+\tilde u)^{-1}(1-{u\over w})\approx 1-\tilde u({\tilde
w+1 \over \tilde w})$.

 In order to have $\beta \to 0$ and $a $ finite, 
we must have $u, v \to 0$, 
\eqn\epS{ \tilde \epsilon \equiv {2\tilde v\over \sqrt{\tilde w}} \to 
0.}
 If one wants to keep, according to \SeONE , $a$ finite
in this limit, $u/v$ must go to zero such that
\eqn\LimE{-  \tilde \epsilon \ln {u\over v} =
2 a_0 }
finite, i.e. $u\to 0$ exponentially faster than $\tilde 
v/\sqrt{\tilde w}$
(and the $\approx$ sign in (B10) and thereafter, means that such terms
are dropped). 

One also finds that 
 ${u\over w}\to 0$ (even if $w$ $\to 0$)
 as if not, the r.h.s. of \Sdve\ would go to zero.
 So $\beta\to 0$, $a\to a_0 >0$ implies  
\eqn\lIMit{
u,v\to 0, \ \ \ {u\over v} \to 0, \ \ {u\over w} \to 0}
together with \epS , \LimE , and
\eqn\LIme{ \beta \approx \tilde \epsilon \pi.}
In order to extract more quantitative information from 
\Sdve\ consider the
equivalent condition \SIXdve , 
\eqn\sixDVE{(\tilde u+\tilde v)
K(k) = \tilde u {\sl \Pi} \left({1\over 1+\tilde u}, k\right)}
where
\eqn\LimK{k^2 =  \left(1-{u\over w} \right)
 \left({1\over 1+ \tilde u} \right) <{1\over 
1+ \tilde u}.}
 As $\tilde u\to 0$, $k'^2 = 1-k^2
\approx {u\over w} - \tilde u \to 0$ and we can use some standard
expansions for the third elliptic integral appearing in \sixDVE , 
e.g.  
( 412.01 of [BF])
\eqn\limEXP{ 
 {\sl \Pi} \left({1\over 1+\tilde u}, k\right)
= K(k) +{\pi\over 2} {1\over \sqrt{1+\tilde u}}
{ 1 - \Lambda_0(\theta, k)\over \sqrt{ {\tilde u\over
1 + \tilde u} { 1 \over 1 + \tilde u}{\tilde u\over 
\tilde w}}}}
where 
\eqn\LImM{\sin\theta = \sqrt{ {\tilde u\over
1 + \tilde u}\over {u\over w} +  {\tilde u\over
1 + \tilde u}}\approx \sqrt{\tilde w\over \tilde w +1}
}
and the first terms in the expansion of Heumann's Lambda function 
$\Lambda_0(\theta, k)$
(904.00 of [BF]) are
\eqn\lamBDAo{
\Lambda_0(\theta, k)=
{2\over \pi} \left(
E\theta - {1\over 4} (2K-E) k'^2(\theta -
\sin\theta \cos\theta ) + ...\right) \approx
{2\over \pi} \arcsin\sqrt{\tilde w \over \tilde w +1}.}
Inserting \LImM\ into \sixDVE\ and using
$$K(k) \approx - \hf \ln k'^2 +\ln 4 \ \ \ {\rm for} \ \ \ 
k'\to 0$$
one finds:
$$ \tilde v \left(-\hf \ln
\left({u\over w} + {\tilde u \over 1+ \tilde u}\right)
+\ln 4\right)
\approx {\pi\over 2}\sqrt{\tilde w}
\left(1-{2\over \pi}\arcsin \sqrt{ \tilde w \over 1+ \tilde 
w}\right),$$
and, using \LIme  
\eqn\Trindst{{u\over w} + {\tilde u \over 1+ \tilde u}
 \approx 16 e^{-{2\pi^2\over \beta}(1-{2\over \pi}
\arcsin \sqrt{\tilde w\over \tilde w+1})}}
\eqn\TdeU{\tilde u \approx  {16 \tilde w \over 1+ \tilde w}
 e^{-{2\pi^2\over \beta}\left(1-{2\over \pi}
\arcsin \sqrt{\tilde w\over \tilde w+1}\right)
}.}

Apart from converting $\ln \tilde u$ terms into 
$\tilde w, \beta$ dependencies, all other $\tilde u$-dependencies are
dropped,  due to this exponential decay \TdeU . Eq. (B.1.5)  \ \ can 
then
be stated explicitly as an expression for $a_{31}$ in terms
of $\beta$ and $\tilde w$ as follows:
\eqn\PPEt{\eqalign{\ln a_5 &\approx \int_0^\infty ds \left({1\over 
\sqrt{ (s+a_{31}+w)s}} - {1\over s+a_3 +w}\right)
+{v\over w} \int_1^\infty {ds\over s\sqrt{(s-1)(\beta +{1\over \tilde 
w})}}\cr 
 &=\lim_{\Lambda\to \infty}
\left(\ln\left(2\sqrt{s^2 +sa_{51}} + 2s 
+a_{51}\right)\Big|_0^\Lambda+ \ln{a_5\over\Lambda} \right)  + 
{v\over w}\sqrt{\tilde w}\left(\arcsin\left({1-\tilde w\over 
1+\tilde w}\right) +{\pi\over 2}\right)
\cr 
&= \ln 4 +\ln {a_5 \over a_{51}} + {\beta\over 2\pi}
\left(\arcsin\left({1-\tilde w\over 
1+\tilde w}\right) +{\pi\over 2}\right)
\cr}} 
Hence
\eqn\ATrE{ a_{31} \approx {4\over 1+\tilde w} \ e^{
{\beta\over 4} \left[ 1+{2\over \pi} 
\arcsin\left({1-\tilde w\over 
1+\tilde w}\right)\right]}.}
The last equation needed to calculate the $a_i$ as functions of 
$\beta\to 0$ and $\lambda$ is \SIXsst , resp.
\eqn\LASTT{{4\pi \sinh(\beta/2)\over \lambda}
=\sqrt{w(a_{31} +u)} E(k) +
{2va_3\over \sqrt{wa_{31}}}K(k)}
as $\gamma^2 = w(a_{31}+u)$ and (due to (B.1.4))
\eqn\LasT{ a_4a_5+a_1a_3 = - 2 v a_3.}
\LASTT\ can be simplified substantially even without neglecting 
$\tilde u$-terms, by noting that \sixDVE ,
(B.17) imply 
\eqn\lAST{{1\over \pi} {2 v\over \sqrt{ wa_{31}}}
K(k) = \sqrt{1+\tilde u} \ [1-\Lambda_0(\theta, k)].}
With 
\eqn\lasT{{a_3\over 2} = {a_{31} \over 4} (1-2\tilde v - 
\tilde w -\tilde u), \ \ \tilde v ={\beta\over 2\pi} \sqrt{\tilde w}}
one therefore gets
\eqn\lASt{\eqalign{{2 \sinh(\beta/2)\over \lambda}
=  &{a_{31} \over 4} \sqrt{1+\tilde u} \left(
{2\over \pi}  \sqrt{ \tilde w} E+
 (1-2\tilde v - 
\tilde w -\tilde u)(1-\Lambda_0)\right)\cr
\approx & {a_{31}\over 4} \left( {2\over \pi}  \sqrt{ \tilde w} + 
\left(1-{\beta\over\pi}  \sqrt{ \tilde w}  
-    \tilde w \right)\left( 1-{2\over \pi} \arcsin
\sqrt{ \tilde w\over   \tilde w+1}\right) \right)\cr}}
which is an (implicit) equation for $\tilde w$ 
as a function of $\beta$ and $\lambda$, when inserting \ATrE . 
For $\tilde w \to 0$ it reads
  \eqn\Pisnami{ {1-e^{-\beta}\over \lambda} \approx
1- {\beta\over \pi} \sqrt{ \tilde w}
-2\tilde w.} 

The length of the cut is given by (cp. (B.1.1),(B9))
\eqn\uraaa{ 
a\to a_0 = \pi \left( 1- {2\over \pi} \arcsin\sqrt{\tilde w\over 
\tilde w +1}\right)
}

The second of the final scaling eqs. \frenSL\ follows from \lASt\ and 
\uraaa\ if we neglect all terms proportional to $\beta$ or 
exponentially small terms and use $a=2v_{\infty}$.

Finally note that the second line of (B.10), via $K(r)\approx {1\over
2}\ln {16\tilde w\over\tilde u(\tilde w+1)}$ implies (B.21)
(shortcutting the argument (B.15--21)), when using
$$ \sqrt{\tilde w}\int_0^1{ds\over\sqrt{(1-s)(1+s\tilde w)}}={\pi\over
2} -\arcsin{1-\tilde w\over 1+\tilde w}$$
and 
$$\eqalign{{1\over 2}+{1\over\pi} \arcsin{1-\tilde w\over 1+\tilde w}
&= {2\over\pi}\arcsin{1\over \sqrt{1+\tilde w}}\cr
&= 1-{2\over\pi}\arcsin \sqrt{\tilde w\over\tilde w+1}.\cr
}$$


\appendix{C}{Inverted oscillator: the point $\beta =i \pi$ }

An interesting analytical continuation of our  model corresponds to
the imaginary values of the generator of $O_\e(2)$ symmetry of the
original supersymmetric model.  If we renormalise $\e$ to one it is
equivalent to the change $\b \ra i\b$ in \twpa. The corresponding
saddle point equation reads:

\eqn\speib{
2\lambda  \sin  u  =\int_{- a}^{a}
\!  \! \! \! \! \! \! \! \! \! -   d u' 
  \rho(u')\left(2\cot {u-u'\over 2} 
 -\cot {u -u'+\beta\over 2}
- \cot{u -u'-\beta\over 2}
   \right)
}

According to the arguments and results of the paper \BUKA\ the
inverted twisted matrix oscillator describes the compactified $c=1$
string, or, in other words, the compactified bosonic field coupled to
the 2d quantum gravity. So at least the critical regime of $c=1$ 
string
with the typical inverse logarithmic dependence of the physical
quantities on the cosmological coupling should show up at some
point. Let us demonstrate it in the case which we can solve
explicitly, namely for $\beta=i\pi$. The equation \speib\ in this case
looks as:

\eqn\Volll{{\la\over 2} \sin u  
  =\int_{-a}^{ a}
\!   \! \! \! \! \! \! \! \! \! -   \ \ du' 
  { \rho(u') 
\over \sin  (u -u')  }.}

The spectral density is 
$$\rho(u) = {\la \over 2\pi} \sqrt{\sin^2{a} - \sin^{2}u}
.$$
The normalization condition gives
$$1=\int_{-a}^{a}du \rho(u)={\la \over 2\pi } 
\int_{-a}^{a}du \sqrt{\sin^2{a} - \sin^{2}u} =  {2\la\over \pi } 
[ E(\sin a) - \cos^{2}a\  K(\sin a) ].
$$
or
\eqn\meq{ E(k) - k'^{2}  K(k) = {\pi \over \la}, 
\ \ \ k = \sin a. }
Consider the limit when the eigenvalues occupy almost the whole
interval $[-\pi,\pi]$ allowed by the periodicity: $a\sim\pi, \ \
 k^2\simeq 1$. In terms of $k'$ we have the following
assymptotics:
\eqn\assE{ E\simeq 1+{1\over 2}k'^2\log(4/k') }
\eqn\assK{ K\simeq  \log(4/k')  }
By the use of \assE and \assK we obtain from   \meq:
\eqn\sclg{ k'^2 \simeq {2\over \pi} {(\la-\la_c)\over |\log(\la-\la_c)|}  }
for $\la\ra \la_c=\pi$.

For the simplest physical quantity: the derivative of the free energy
we obtain:
\eqn\fgpr{     F'_\la =
 \int_{-a}^a du \rho(u) \cos u
={\la k^2 \over 2}} 
from where we obtain the scaling asymptotics typical for the
$c=1$ noncritical string discovered in \KM :
\eqn\cosc{ F(\la) \simeq {\pi^2\over 4} -
{1\over  4} {(\la-\la_c)^2\over |\log(\la-\la_c)|}  }

The considered case $\beta=i\pi$ of the $c=1$ matrix model corresponds
to the Kosterlitz-Thouless phase transition point. It would be
interesting to study the vicinity of this point by generalizing our
solution to all imaginary $\beta$.

\listrefs 
\bye